\documentclass[11pt]{article}

\usepackage[]{acl}

\usepackage{times}
\usepackage{latexsym}

\usepackage[T1]{fontenc}

\usepackage[utf8]{inputenc}

\usepackage{microtype}

\usepackage{inconsolata}

\usepackage{graphicx}

\usepackage{titletoc} 
\usepackage{hyperref} 

\usepackage{times}
\usepackage{latexsym}
\usepackage{multirow}
\usepackage{listings}
\usepackage{pifont}

\usepackage{graphicx}
\usepackage{amsmath}
\usepackage{wrapfig}
\usepackage{booktabs}
\usepackage{xcolor,colortbl}
\usepackage{tcolorbox}
\tcbuselibrary{skins, breakable}
\usepackage{subcaption}
\usepackage{graphicx}
\newcommand{\graybg}{\rowcolor{gray!20}}
\newcommand{\lightgraybg}{\rowcolor{gray!10}}

\usepackage{fontawesome5}

\newcommand{\upred}[1]{\textcolor{purple}{\tiny \,$\uparrow$#1}}
\newcommand{\downblue}[1]{\textcolor{blue}{\tiny \,$\downarrow$#1}}
\usepackage{tcolorbox} 
\definecolor{skyblue}{RGB}{203, 221, 245}
\newcommand{\skyblue}{\rowcolor{skyblue}}

\newtcbox{\roundbox}{on line, 
    colback=gray!10,      
    colframe=gray!50,     
    boxrule=0.5pt,        
    arc=2pt,              
    outer arc=2pt,        
    boxsep=1pt,           
    left=1pt, right=1pt, top=0.5pt, bottom=0.5pt 
}

\title{From Completion to Editing: Unlocking Context-Aware Code Infilling via Search-and-Replace Instruction Tuning}

\author{Jiajun Zhang\textsuperscript{1, 3},  Zeyu Cui\textsuperscript{2}, Jiaxi Yang\textsuperscript{4}, Lei Zhang\textsuperscript{4}, \textbf{Yuheng Jing\textsuperscript{3}}, \textbf{Zeyao Ma\textsuperscript{2}} \\
 \textbf{Tianyi Bai\textsuperscript{2}}, \textbf{Binyuan Hui\textsuperscript{2}}, \textbf{Qiang Liu\textsuperscript{3}}, \textbf{Zilei Wang\textsuperscript{1}}, \textbf{Liang Wang\textsuperscript{3}}, \textbf{Junyang Lin\textsuperscript{2}} \\
 \textsuperscript{1} USTC \quad 
 \textsuperscript{2} Alibaba Group. \quad
 \textsuperscript{3} CASIA \quad
 \textsuperscript{4} SIAT \quad \\[0.2em]
 {\faEnvelope} \texttt{zhangjiajun519@gmail.com} \\[0.2em]
 {\faGlobe} \textbf{Project Page:} \href{https://github.com/QwenLM/Qwen3-Coder}{Qwen3-Coder} \\[0.2em]
}

\begin{document}

\maketitle
\thispagestyle{firstpage}

\begin{abstract}
The dominant Fill-in-the-Middle (FIM) paradigm for code completion is constrained by its rigid inability to correct contextual errors and reliance on unaligned, insecure Base models. While Chat LLMs offer safety and Agentic workflows provide flexibility, they suffer from performance degradation and prohibitive latency, respectively. To resolve this dilemma, we propose \underline{\textbf{S}}earch-and-\underline{\textbf{R}}eplace \underline{\textbf{I}}nfilling (\textbf{SRI}), a framework that internalizes the agentic verification-and-editing mechanism into a unified, single-pass inference process. By structurally grounding edits via an explicit search phase, SRI harmonizes completion tasks with the instruction-following priors of Chat LLMs, extending the paradigm from static infilling to dynamic context-aware editing. We synthesize a high-quality dataset, \textbf{SRI-200K}, and fine-tune the \textbf{SRI-Coder} series. Extensive evaluations demonstrate that with minimal data (20k samples), SRI-Coder enables Chat models to surpass the completion performance of their Base counterparts. Crucially, unlike FIM-style tuning, SRI preserves general coding competencies and maintains inference latency comparable to standard FIM. We empower the entire Qwen3-Coder series with SRI, encouraging the developer community to leverage this framework for advanced auto-completion and assisted development.
\end{abstract}
\section{Introduction}
Large Language Models (LLMs) have revolutionized software development, with the Fill-in-the-Middle (FIM) paradigm serving as the industry standard for code completion. By conditioning on both prefix and suffix contexts, FIM enables models to generate missing code segments effectively \citep{fim}. The FIM strategy has significantly advanced the code completion capabilities of numerous LLMs, including CodeX \cite{codex}, StarCoder \cite{starcoder2}, DeepSeekCoder \cite{deepseek_coder}, Qwen-Coder \cite{qwen25coder}, and CodeLlama \cite{code_llama}. Leveraging these models, LLM-powered programming tools such as \textit{GitHub Copilot}\footnote{\href{https://github.com/features/copilot}{https://github.com/features/copilot}},  \textit{Augment}\footnote{\href{https://docs.augmentcode.com/}{https://docs.augmentcode.com}} and \textit{Continue.dev}\footnote{\href{https://www.continue.dev/}{https://www.continue.dev}} have implemented powerful autocomplete features, which leverages context-aware suggestions and a convenient "\textit{Tab-to-complete}" interface to boost developer productivity.

\begin{figure}
    \centering
    \includegraphics[width=\linewidth]{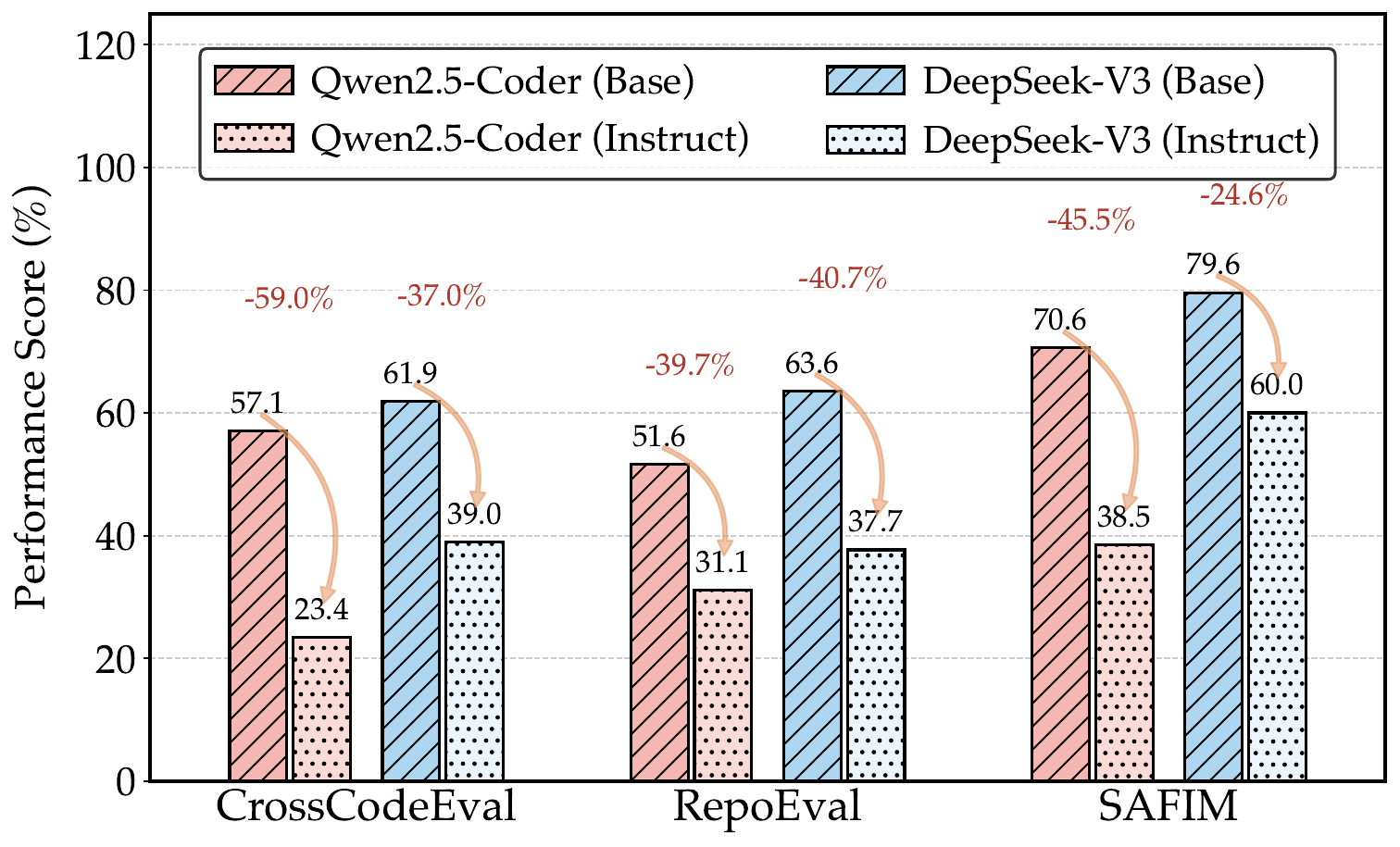}
    \caption{Performance disparity between Base models and their Instruction-tuned counterparts on standard code completion benchmarks. The results highlight a significant degradation in completion capabilities (up to 59.0\%) when prompting safety-aligned Chat LLMs for FIM tasks, demonstrating that FIM skills do not naturally transfer to the conversational setting.}
    \label{fig:base_chat_compare}
    \vspace{-1em}
\end{figure}

However, this paradigm exhibits two critical drawbacks: \ding{182} \textbf{The Optimal Context Assumption:} FIM operates on the unrealistic premise that the surrounding context is error-free. Consequently, when the context contains bugs, FIM is incapable of correcting them; it merely builds upon the flaws, inevitably resulting in broken code. \ding{183} \textbf{Inherent Security Risks:} FIM typically relies on unaligned base LLMs, which lack the safety alignments of chat LLMs. This exposes systems to severe vulnerabilities; notably, the SAL benchmark demonstrates that FIM-based tools like \textit{GitHub Copilot} suffer from attack success rates exceeding 99\% \citep{cheng2025security}. While directly prompting safety-aligned Chat LLMs to perform FIM tasks might seem like a solution, Figure \ref{fig:base_chat_compare} illustrates that this approach causes significant performance degradation, failing to effectively transfer completion capabilities to the instruction-tuned setting.

The conventional FIM paradigm is constrained by inherent rigidity and security vulnerabilities, while directly utilizing Chat LLMs for infilling often results in performance degradation. Furthermore, although multi-step agentic workflows offer enhanced capabilities, they incur latency overheads incompatible with real-time completion scenarios. To resolve this dilemma, we propose \underline{\textbf{S}}earch-and-\underline{\textbf{R}}eplace \underline{\textbf{I}}nfilling (\textbf{SRI}), a framework that internalizes the agentic verification-and-editing mechanism into a unified, single-pass inference process. By structurally grounding edits through an explicit search phase, SRI not only harmonizes completion tasks with the instruction-following priors of Chat LLMs but also extends the paradigm from static infilling to dynamic context-aware editing. This end-to-end approach synthesizes the high performance and flexibility of agentic editing with the computational efficiency essential for interactive development.

Methodologically, we first substantiate the intrinsic advantages of SRI in mitigating security vulnerabilities and resolving contextual inconsistencies through controlled experiments (\S\ref{sec:sri_method}). Building on this foundation, we synthesize \textbf{SRI-200K}, a high-quality dataset derived from \textit{The Stack v2} \citep{starcoder2}, and construct the \textbf{SRI-Coder} series by fine-tuning the Qwen2.5-Coder family. We then conduct comprehensive evaluations across over 20 LLMs on both mainstream code completion and general coding benchmarks. Our empirical results yield four critical findings: \ding{182} SRI consistently and significantly outperforms various natural language FIM prompting strategies across Chat LLMs (\S\ref{sec:fim_sri_compare}); \ding{183} with minimal fine-tuning (20k samples), SRI-Coder enables Chat models to surpass the completion performance of their corresponding Base counterparts (\S\ref{sec:main_result}); \ding{184} unlike FIM-style fine-tuning, which degrades general coding capabilities, SRI acts as a superior tuning format that enhances completion skills while preserving the model's broader competencies (\S\ref{sec:fim_sri_compare}); and \ding{185} crucially, SRI maintains inference latency comparable to standard FIM, validating its feasibility for real-time deployment despite the structured generation process (\S\ref{sec:scale_study}). Collectively, these findings establish SRI as a robust and efficient alternative for next-generation auto-completion tools, offering a promising direction for aligning large language models with interactive development workflows.


\section{Preliminary}

\begin{figure*}
    \centering
    \includegraphics[width=\linewidth]{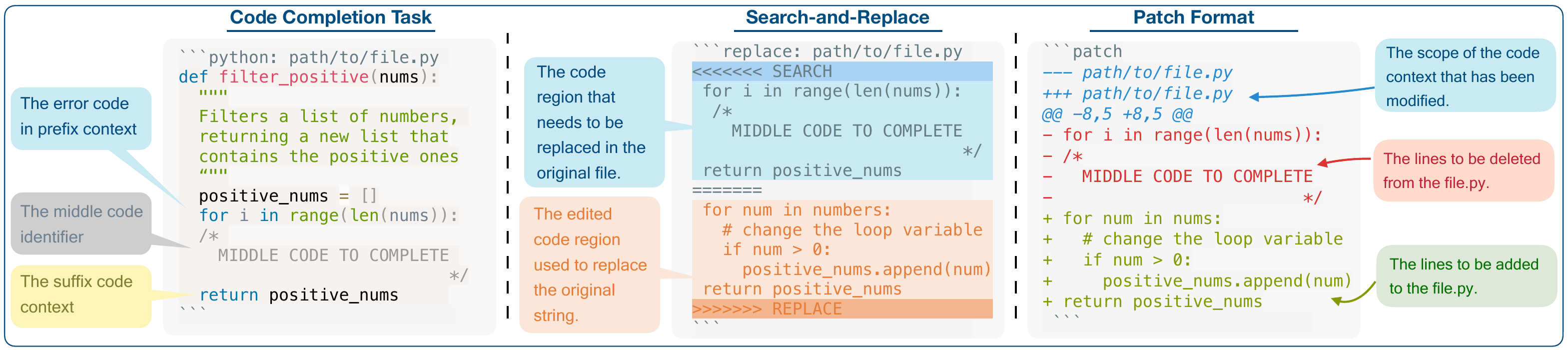}
    \caption{The workflow of our proposed Search-and-Replace Infilling (SRI) method. (1) The process begins with a code completion task containing a pre-existing bug (an incorrect loop variable) and an \roundbox{\texttt{/* MIDDLE CODE TO COMPLETE */}} identifier marking the target location. (2) Our SRI model generates a search-and-replace block that simultaneously infills the missing function body and corrects the contextual bug. (3) This block can then be converted into a standard patch format for versatile application.}
    \vspace{-1.5em}
    \label{fig:sri_workflow}
\end{figure*}

\textbf{Fill-in-the-middle Code Completion}. Fill-in-the-middle (FIM) is a pre-training strategy and inference format designed for base large language models (LLMs). It has been widely adopted in code-centric LLMs such as Qwen-Coder \cite{qwen25coder,deepseek_coder,starcoder2} and even some general-purpose models like DeepSeek-V3 \cite{deepseekv3}. FIM addresses a core limitation of decoder-only architectures—their inability to perform infilling tasks—through a data transformation technique applied during both training and inference. This allows the model to leverage both prefix and suffix contexts to generate an intermediate code segment.

During FIM pre-training, each document is split into three parts: \texttt{prefix}, \texttt{middle}, and \texttt{suffix}. These components are independently encoded and marked with special sentinel tokens \roundbox{\texttt{<PRE>}}, \roundbox{\texttt{<MID>}}, and \roundbox{\texttt{<SUF>}}, respectively. The final tokenized sequence follows a specific order: \texttt{prefix} $\rightarrow$ \texttt{suffix} $\rightarrow$ \texttt{middle}, with their corresponding sentinel tokens preserved, as shown in Equation \ref{equa:train}:
\vspace{-5pt}
\begin{equation}
    \label{equa:train}
    \text{\roundbox{\texttt{<PRE>}}} \circ \text{\texttt{prefix}} \circ \text{\roundbox{\texttt{<SUF>}}} \circ \text{\texttt{suffix}} \circ \text{\roundbox{\texttt{<MID>}}} \circ \text{\texttt{middle}}
\end{equation}

where $\circ$ denotes concatenation. During inference, given a prefix and suffix pair, we construct the input prompt by removing the middle segment and its content while maintaining the special tokens, as expressed in Equation \ref{euqa:infer}:
\vspace{-5pt}
\begin{equation}
    \label{euqa:infer}
    \text{\roundbox{\texttt{<PRE>}}} \circ \text{\texttt{prefix}} \circ \text{\roundbox{\texttt{<SUF>}}} \circ \text{\texttt{suffix}} \circ \text{\roundbox{\texttt{<MID>}}}
\end{equation}


This structured prompt guides the model to synthesize a coherent middle segment by conditioning on both the prefix and suffix. By effectively resolving the infilling challenge for decoder-only LLMs, FIM has been instrumental in the advancement of LLM-powered autocomplete coding tools.

\label{sec:fim_chat}
\paragraph{Code Completion in Chat LLMs}
While Base LLMs perform FIM via raw text continuation, Chat LLMs are constrained by post-training alignment formats (e.g., \texttt{ChatML} \cite{chatml}). The mandatory application of chat templates (\roundbox{\texttt{apply\_chat\_template}}) structurally conflicts with the special tokens required for standard FIM inference, rendering direct application infeasible.

To bypass this limitation, practitioners typically employ natural language (NL) prompts to simulate FIM tasks. Common adaptations include \textit{Standard FIM Completion} (Figure \ref{fig:prompt_standard}), \textit{Dialogue-Based Reconstruction} (Figure \ref{fig:prompt_dialogue}), and \textit{Template-Based Infilling} (Figure \ref{fig:prompt_template}). However, the completion proficiency acquired during pre-training fails to effectively transfer to these conversational settings, resulting in significant performance degradation compared to Base models. A detailed theoretical analysis in Appendix \ref{sec:theory} further elucidates the inherent limitations of relying on unstructured NL instructions for precise code generation.

\paragraph{The Search-and-Replace Format}
The \texttt{diff} format, used to represent the differences between two files, is widely employed in applications such as version control (e.g., Git commits), code comparison tools, and coding agents. A common implementation is the Unified Diff Format, often referred to as a \texttt{patch}. A patch is a text file that declaratively specifies the changes required to transform a file from one state to another (e.g., "\textit{delete these lines, add those lines}"). It precisely identifies changes based on line numbers and context.
However, generating a syntactically correct patch is a significant challenge for LLMs, as they often struggle to produce accurate line numbers \cite{swe-flow}. To address this limitation, \citet{swe-rl} proposed a Search-and-Replace format as a more robust alternative.
This format, inspired by the GitHub merge conflict convention, uses explicit \texttt{SEARCH} and \texttt{REPLACE} blocks to delineate the code to be modified and its replacement. This structure effectively captures code edits by clearly defining the \texttt{before} and \texttt{after} states without relying on fragile line numbers. Its clear structure and high readability make it highly accessible to both humans and LLMs, leading to its widespread adoption in LLM-based software engineering tasks \cite{agentless,swe-bench,openhands}.
\section{From Static Completion to Context-Aware Editing}
\label{sec:sri_method}

\begin{figure}[t]
    \centering
    \includegraphics[width=0.95\linewidth]{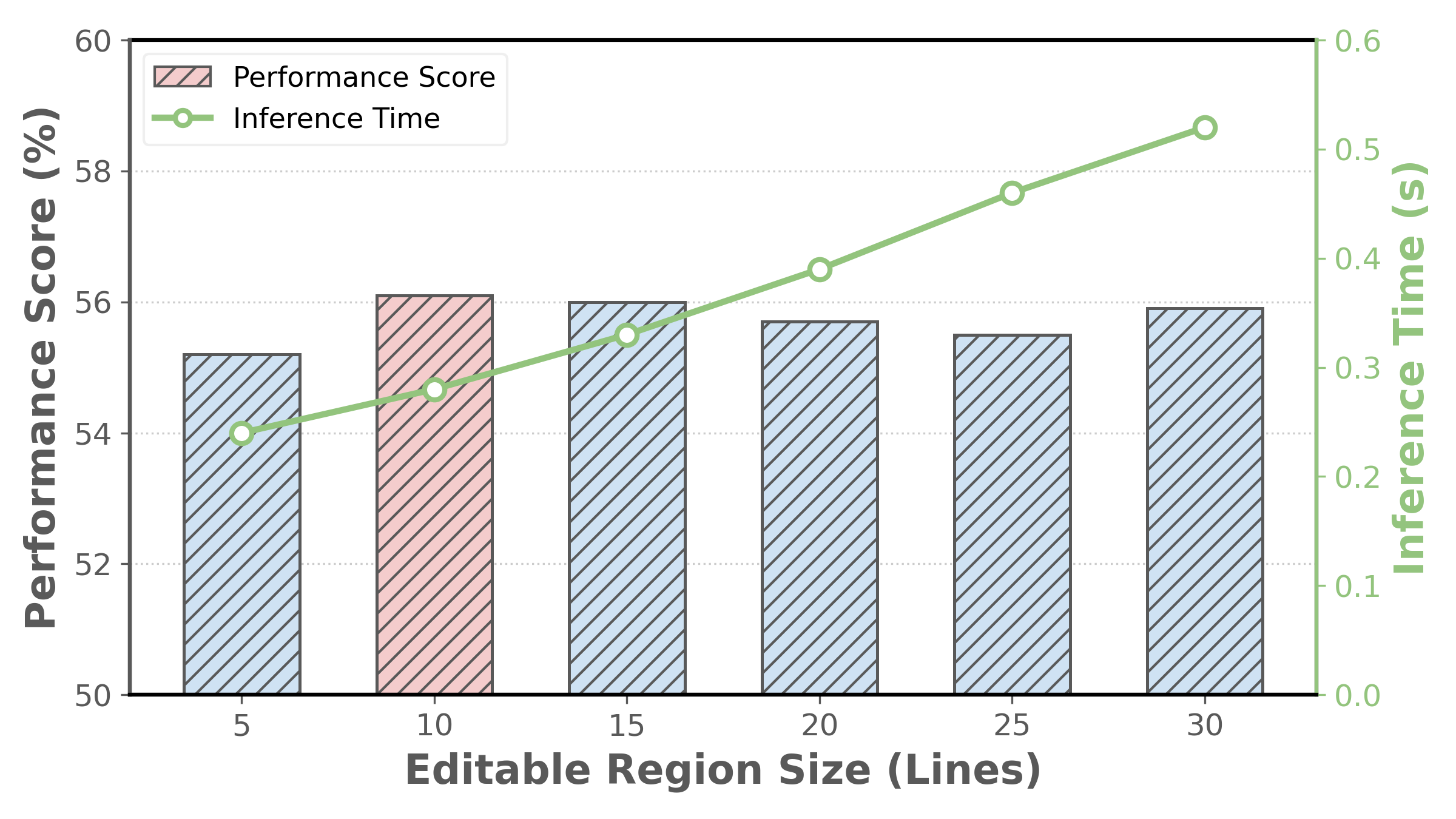}
    \caption{SRI editable region size sensitivity analysis.}
    \label{fig:line_sensitivity}
    \vspace{-1em}
\end{figure}

In this section, we introduce \textbf{Search-and-Replace Infilling (SRI)}, a paradigm shift that transitions code completion from static text continuation to intelligent micro-editing. We structure our discussion as follows: Section \ref{sec:sri_method} details the grounding mechanism behind SRI, Section \ref{sec:enhanced_security} demonstrates how SRI bridges the safety gap, and Section \ref{sec:enhanced_flexibility} showcases its robustness to contextual noise.

\subsection{Methodology: Contextual Grounding}

To address the limitations of static FIM infilling, we propose \underline{\textbf{S}}earch-and-\underline{\textbf{R}}eplace \underline{\textbf{I}}nfilling (SRI). Unlike FIM, which generates the middle segment without structural validation, SRI reframes the task as \textbf{context-aware micro-editing}. The core innovation lies in its workflow, as illustrated in Figure \ref{fig:sri_workflow}.
In the SRI framework, we designate the target location with a unique identifier, \roundbox{\texttt{/* MIDDLE CODE TO COMPLETE */}}. Functionally analogous to the cursor in standard auto-completion tools, this marker explicitly denotes the active position for generation. We then define an \textit{editable region} comprising the identifier and 10 lines of surrounding context (Figure \ref{fig:line_sensitivity}). Crucially, rather than immediately generating the new code, the model is prompted to first generate a \texttt{SEARCH} block that explicitly replicates the code segment intended for modification.

This \texttt{SEARCH} generation functions as a \textbf{grounding mechanism}, operating akin to a visual Chain-of-Thought. It compels the model to identify and validate the context before committing to any changes. Following this verification step, the model generates the \texttt{REPLACE} block. This structured output allows for direct application via IDE tools or conversion into standard patch formats (e.g., \texttt{git apply}).
We select the search-and-replace format over standard diff patches for its \textbf{robustness}. While standard patches rely on fragile line numbers (which LLMs struggle to count accurately \cite{swe-flow}), the search-and-replace format anchors edits based on semantic content. This ensures reliable execution even when the model's internal line counting is imprecise.

\vspace{-0.5em}

\subsection{Bridging the Security Alignment Gap}
\label{sec:enhanced_security}
The traditional FIM paradigm forces the use of unaligned Base models, leaving systems inherently vulnerable to code injection. SRI addresses this critical flaw by serving as an \textbf{alignment bridge}, enabling code completion tasks to utilize instruction-tuned Chat models and effectively \textbf{inherit} their robust post-training safety features.

To validate this, we utilize the SAL benchmark \cite{cheng2025security} to evaluate resistance against injection attacks. We compare the vulnerability of state-of-the-art Base models (using FIM) against their aligned Instruct counterparts (using SRI). The results for Level 1 (direct) and Level 2 (complex) attacks are presented in Table \ref{tab:sal_results_en}, with additional benchmark results detailed in Appendix \S\ref{app-sec:detail_result}.
The data reveals a fundamental disparity: Base models are nearly defenseless, exhibiting attack success rates approaching 100\%. In contrast, SRI-enabled models leverage the safety alignment of their Chat backbones to drastically reduce vulnerability. Notably, this architecture also allows for the integration of \textit{Safe Prompts} (unlike FIM), which can further suppress successful attacks to near-zero levels.

\begin{table}[t]
  \centering
  \small
  \begin{tabular}{l c c}
    \toprule
    \textbf{Model} & \multicolumn{2}{c}{\textbf{Attack Success Rate (\%)}} \\
    \cmidrule(lr){2-3} 
  \skyblue{}  & \textbf{Level 1} & \textbf{Level 2} \\
    \midrule
    \multicolumn{3}{l}{\textit{Commercial Baseline}} \\
    \hspace{1em}GitHub Copilot & 98.8 & 99.4 \\
    \midrule
    \multicolumn{3}{l}{\textit{Qwen2.5-Coder Series}} \\
    \hspace{1em}Qwen-Base (FIM)         & 100.0 & 97.5 \\
    \hspace{1em}Qwen-Instruct (SRI)       & \textbf{13.8}  & \textbf{20.0} \\
    \midrule
    \multicolumn{3}{l}{\textit{DeepSeek-V3 Series}} \\
    \hspace{1em}DeepSeek-Base (FIM)     & 98.2  & 95.4 \\
    \hspace{1em}DeepSeek-Instruct (SRI) & \textbf{5.5}  & \textbf{7.1} \\
    \bottomrule
  \end{tabular}
    \caption{Security evaluation on the SAL Benchmark. The table compares commercial/base FIM models against aligned Instruct models. Even leading commercial tools like Copilot are vulnerable, whereas SRI effectively bridges the alignment gap.}
    \vspace{-1.5em}
    \label{tab:sal_results_en}
\end{table}

\subsection{Robustness to Contextual Noise}
\label{sec:enhanced_flexibility}
Real-world development environments are rarely perfect; they often contain contextual inconsistencies, such as syntax errors, legacy patterns, or temporary placeholders. Traditional FIM operates under a rigid assumption of frozen context, merely appending code without addressing existing flaws. In contrast, SRI exhibits \textbf{intent-aware robustness}, enabling the model to dynamically rectify the surrounding context to match the user's objective.
To measure this capability, we construct \textbf{CrossCodeEval-Flex}. This benchmark adapts the standard CrossCodeEval dataset \cite{crosscodeeval} by simulating common developer error patterns. Specifically, we introduce controlled syntactical or logical noise into the 5-line window surrounding the missing middle code. This setup creates a challenging scenario where the model must perform simultaneous completion and correction.
The results in Table \ref{tab:cceval_f_results_en} highlight a critical disparity across diverse model families. Standard FIM models, including both \textbf{Qwen2.5-Coder-32B} and the state-of-the-art \textbf{DeepSeek-V3-Base}, fail completely and yield near 0.0 EM scores as they are architecturally constrained to preserve the noisy context. In contrast, the Chat LLMs utilizing SRI successfully identify and resolve the inconsistencies. Specifically, \textbf{SRI-Coder-32B}, the model we fine-tuned for this paradigm, achieves the most significant improvements. This confirms that SRI transcends the limitations of static completion, offering a resilient solution for complex editing scenarios regardless of the underlying model architecture.

\begin{table}[t]
  \centering
  \small
  \begin{tabular}{l c c}
    \toprule
    \textbf{Model} & \multicolumn{2}{c}{\textbf{CrossCodeEval-Flex}} \\
    \cmidrule(lr){2-3} 
 \skyblue{}    & \textbf{EM} & \textbf{ES} \\
    \midrule 
    \multicolumn{3}{l}{\textit{Qwen2.5-Coder Series}} \\
    \hspace{1em}Qwen-Base (FIM)           & 0.0 & 35.6 \\
    \hspace{1em}Qwen-Instruct (SRI)       & 14.0 & 53.8 \\
    \hspace{1em}\textbf{SRI-Coder}         & \textbf{33.0} & \textbf{71.5} \\
    \midrule
    \multicolumn{3}{l}{\textit{DeepSeek-V3 Series}} \\
    \hspace{1em}DeepSeek-Base (FIM)       & 0.0 & 38.2 \\ 
    \hspace{1em}DeepSeek-Chat (SRI)       & 24.5 & 67.4 \\ 
    \bottomrule
  \end{tabular}
    \caption{Performance on CrossCodeEval-Flex. We evaluate both the Qwen2.5-Coder-32B series and the DeepSeek-V3 series. Standard FIM fails completely in noisy environments (0.0 EM), whereas SRI successfully acts as a corrective editor. Note that SRI-Coder refers to the model fine-tuned on our synthesized dataset.}
    \vspace{-10pt}
    \label{tab:cceval_f_results_en}
\end{table}
\begin{table*}[h]
\centering
\small
\setlength{\tabcolsep}{2pt}
\begin{tabular}{@{}llcccccccccccc@{}}
\toprule
& \multirow{3}{*}{\textbf{Model}} & \multicolumn{4}{c}{\textbf{CrossCodeEval}} & \multicolumn{4}{c}{\textbf{RepoEval}} & \multicolumn{4}{c}{\textbf{CrossCodeLongEval}} \\
\cmidrule(lr){3-6} \cmidrule(lr){7-10} \cmidrule(lr){11-14}
& & \multicolumn{2}{c}{\textit{EM}} & \multicolumn{2}{c}{\textit{ES}} & \multicolumn{2}{c}{\textit{EM}} & \multicolumn{2}{c}{\textit{ES}} & \multicolumn{2}{c}{\textit{EM}} & \multicolumn{2}{c}{\textit{ES}} \\
\cmidrule(lr){3-4} \cmidrule(lr){5-6} \cmidrule(lr){7-8} \cmidrule(lr){9-10} \cmidrule(lr){11-12} \cmidrule(lr){13-14}
& & \textbf{FIM} & \textbf{SRI} & \textbf{FIM} & \textbf{SRI} & \textbf{FIM} & \textbf{SRI} & \textbf{FIM} & \textbf{SRI} & \textbf{FIM} & \textbf{SRI} & \textbf{FIM} & \textbf{SRI} \\
\midrule
& \multicolumn{13}{c}{\textbf{Base Models}}\\
\midrule
\lightgraybg& Qwen2.5-Coder-32B & 57.1 & - & 86.8 & - & 51.6 & - & 78.5 & - & 36.8 & - & 66.4 & - \\
& DeepSeek-V3-Base & 61.9 & - & 88.5 & - & 63.6 & - & 86.2 & - & 50.5 & - & 77.9 & - \\
\midrule
& \multicolumn{13}{c}{\textbf{Non-reasoning Chat Models}}\\
\midrule
\lightgraybg& DeepSeek-V3 & 39.0 & 56.1\upred{17.1} & 70.3 & 86.2\upred{15.9} & 37.7 & 54.3\upred{16.6} & 60.4 & 78.9\upred{18.5} & 25.4 & 34.8\upred{9.4} & 53.5 & 67.4\upred{13.9} \\
& DeepSeek-V3-0324 & 44.5 & 56.0\upred{11.5} & 80.0 & 85.7\upred{5.7} & 45.8 & 54.5\upred{8.7} & 68.2 & {79.3}\upred{11.1} & 30.9 & 34.8\upred{3.9} & 60.2 & 68.2\upred{8.0} \\
\lightgraybg& Claude-3.5-Haiku & 23.9 & 44.5\upred{20.6} & 73.2 & 75.2\upred{2.0} & 31.9 & 44.4\upred{12.5} & 58.2 & 67.8\upred{9.6} & 17.7 & 27.1\upred{9.4} & 49.8 & 56.2\upred{6.4} \\
& Claude-3.5-Sonnet & 47.5 & 56.0\upred{8.5} & 82.3 & 85.6\upred{3.3} & 48.0 & 54.4\upred{6.4} & 72.1 & 80.2\upred{8.1} & 32.4 & 36.3\upred{3.9} & 63.2 & 69.4\upred{6.2} \\
\lightgraybg& Claude-3.7-Sonnet & 48.3 & 56.7\upred{8.4} & 81.6 & 85.7\upred{4.1} & 49.3 & 55.0\upred{5.7} & 73.6 & 80.2\upred{6.6} & 32.5 & 38.2\upred{5.7} & 64.0 & \underline{70.5}\upred{6.5} \\
& Claude-4-Sonnet & 57.5 & \underline{60.8}\upred{3.3} & \underline{87.3} & 87.2\downblue{0.1} & 55.7 & \underline{58.9}\upred{3.2} & 78.7 & \underline{82.0}\upred{3.3} & 38.1 & \underline{38.9}\upred{0.7} & 68.8 & 69.5\upred{0.7} \\
\lightgraybg& GPT-4o-1120 & 34.2 & 44.1\upred{9.9} & 72.5 & 80.9\upred{8.4} & 25.7 & 45.0\upred{19.3} & 52.1 & 72.6\upred{20.5} & 21.7 & 23.6\upred{1.9} & 51.6 & 56.4\upred{4.8} \\
& GPT-4o-0806 & 40.3 & 41.7\upred{1.4} & 77.3 & 79.5\upred{2.2} & 37.2 & 46.8\upred{9.6} & 61.5 & 73.7\upred{12.2} & 20.6 & 26.4\upred{5.8} & 50.7 & 58.2\upred{7.5} \\

\lightgraybg& GPT-4o-mini & 9.0 & 27.7\upred{18.7} & 59.6 & 68.3\upred{8.7} & 19.6 & 27.7\upred{8.1} & 48.0 & 54.6\upred{6.6} & 12.1 & 11.1\downblue{1.0} & 41.3 & 39.6\downblue{1.7} \\
& GPT-4.1 & 46.1 & 45.8\downblue{0.3} & 81.0 & 79.8\downblue{1.2} & 47.0 & 46.9\downblue{0.1} & 71.7 & 73.2\upred{1.5} & 28.5 & 30.8\upred{2.3} & 61.2 & 62.2\upred{1.0} \\

\lightgraybg& Qwen3-Coder-480B-A35B & 56.1 & \textbf{62.9}\upred{6.9}  & 85.8 & \textbf{89.5}\upred{3.7} & 51.1 & \textbf{65.1}\upred{14.0} & 73.2 & \textbf{85.4}\upred{12.2} & 36.2 & \textbf{45.9} \upred{9.7} & 65.3 & \textbf{74.7}\upred{9.4} \\
& Qwen2.5-Coder-32B-Instruct & 23.4 & 44.4\upred{21.0} & 69.9 & 76.8\upred{6.9} & 31.1 & 44.8\upred{13.7} & 56.1 & 69.7\upred{13.6} & 20.4 & 25.9\upred{5.5} & 50.3 & 55.8\upred{5.5} \\
\graybg& \textbf{SRI-Coder-32B} & 11.3 & 57.6\upred{46.3} & 37.9 & \underline{87.3}\upred{49.3} & 17.3 & 54.4\upred{37.1} & 37.3 & 78.9\upred{41.5} & 14.6 & 37.1\upred{22.5} & 33.2 & 67.9\upred{34.7} \\
\midrule
& \multicolumn{13}{c}{\textbf{Reasoning Chat Models}}\\
\midrule
\lightgraybg& DeepSeek-R1-0528 & 44.6 & 51.1\upred{6.5} & 82.3 & 84.0\upred{1.7} & 48.1 & 49.2\upred{1.1} & 75.1 & 76.0\upred{0.9} & 29.2 & 29.7\upred{0.5} & 62.2 & 62.6\upred{0.4} \\
& Claude-3.7-Sonnet-think & 49.4 & \textbf{54.6}\upred{5.2} & 82.5 & \textbf{84.5}\upred{2.0} & 9.1 & \underline{53.0}\upred{43.9} & 27.6 & \underline{78.0}\upred{50.4} & 30.6 & \underline{34.9}\upred{4.3} & 60.7 & \underline{66.9}\upred{6.2} \\
\lightgraybg& Grok-3 & 46.4 & \underline{54.4}\upred{8.0} & 82.1 & \underline{84.1}\upred{2.0} & 50.1 & \textbf{53.6}\upred{3.5} & 75.8 & \textbf{78.7}\upred{2.9} & 32.4 & \textbf{35.2}\upred{2.8} & 65.2 & \textbf{68.0}\upred{2.8} \\
& o1-2024-12-17 & 26.9 & 42.7\upred{15.8} & 75.4 & 78.4\upred{3.0} & 44.5 & 48.3\upred{3.8} & 68.4 & 73.6\upred{5.2} & 26.6 & 24.4\downblue{2.2} & 56.6 & 57.0\upred{0.4} \\
\lightgraybg& o3-mini & 23.7 & 27.6\upred{3.9} & 39.1 & 69.9\upred{30.8} & 26.4 & 39.1\upred{12.7} & 37.1 & 61.4\upred{24.3} & 9.7 & 20.9\upred{11.2} & 21.1 & 48.9\upred{27.8} \\
& Qwen3-30B-A3B & 13.7 & 36.1\upred{22.4} & 59.7 & 71.5\upred{11.8} & 20.0 & 33.6\upred{13.6} & 44.3 & 61.2\upred{16.9} & 12.9 & 17.6\upred{4.7} & 36.7 & 45.1\upred{8.4} \\
\lightgraybg& Qwen3-235B-A22B & 14.6  & 42.7\upred{28.2} & 35.6 & 76.8\upred{41.3} & 19.1 & 42.7\upred{23.6} & 30.6 & 68.9\upred{38.3} & 15.0 & 23.2 \upred{8.2} & 32.5 & 53.5\upred{21.0} \\

\bottomrule
\end{tabular}
\caption{Performance comparison on similarity-based benchmarks (CrossCodeEval, RepoEval, and CrossCodeLongEval). For base models, \textbf{FIM} denotes special token based fill-in-the-middle code completion, for chat models, \textbf{FIM} denotes natural language-based code completion prompting, while \textbf{SRI} represents our method. \textbf{Bold} and \underline{underlined} values indicate the best and second-best performance respectively within the model type. \upred{n} and \downblue{n} indicate performance improvements and degradations of SRI compared to FIM.}
\label{tab:ES}
\vspace{-1.5em}
\end{table*}
\begin{table*}[h]
\centering
\small
\setlength{\tabcolsep}{2pt}
\begin{tabular}{@{}llcccccccccccc@{}}
\toprule
& \multirow{3}{*}{\textbf{Model}} & \multicolumn{2}{c}{\textbf{ExecRepoBench}} & \multicolumn{10}{c}{\textbf{SAFIM}} \\
\cmidrule(lr){3-4} \cmidrule(lr){5-14}
& & \multicolumn{2}{c}{\textit{Pass@1}} & \multicolumn{2}{c}{\textit{Block}} & \multicolumn{2}{c}{\textit{Block V2}} & \multicolumn{2}{c}{\textit{API}} & \multicolumn{2}{c}{\textit{Control}} & \multicolumn{2}{c}{\textit{Average}} \\
\cmidrule(lr){3-4} \cmidrule(lr){5-6} \cmidrule(lr){7-8} \cmidrule(lr){9-10} \cmidrule(lr){11-12} \cmidrule(lr){13-14}
& & \textbf{FIM} & \textbf{SRI} & \textbf{FIM} & \textbf{SRI} & \textbf{FIM} & \textbf{SRI} & \textbf{FIM} & \textbf{SRI} & \textbf{FIM} & \textbf{SRI} & \textbf{FIM} & \textbf{SRI} \\
\midrule
& \multicolumn{13}{c}{\textbf{Base Models}}\\
\midrule
\lightgraybg& Qwen2.5-Coder-32B & 25.7    & - & 62.9  & - & 65.2   & - & 75.5   & - & 76.4  & - & 70.0  & - \\
& DeepSeek-V3-Base & 36.5  & - & 76.3  & - & 77.9  & - & 79.4  & - & 84.8  & - & 79.6  & - \\
\midrule
& \multicolumn{13}{c}{\textbf{Non-reasoning Chat Models}}\\
\midrule
\lightgraybg& DeepSeek-V3 & 35.6 & 61.8\upred{26.2} & 60.5 & \textbf{70.1}\upred{9.6} & 59.7 & 70.6\upred{10.9} & 55.8 & 73.6\upred{17.8} & 64.1 & 80.6\upred{16.5} & 60.0 & 73.7\upred{13.7} \\
& DeepSeek-v3-0324 & 36.9 & 65.5\upred{28.6} & 53.3 & \underline{69.8}\upred{16.5} & 56.5 & \underline{71.1}\upred{14.6} & 68.1 & 74.5\upred{6.4} & 65.3 & \textbf{81.8}\upred{16.5} & 60.8 & \underline{74.3}\upred{13.5} \\
\lightgraybg& Claude-3.5-Haiku & 35.6 & 57.7\upred{22.1} & 60.1 & 66.3\upred{6.2} & 59.7 & 68.0\upred{8.3} & 65.2 & 71.6\upred{6.4} & 68.5 & 77.4\upred{8.9} & 63.4 & 70.8\upred{7.4} \\
& Claude-3.5-Sonnet & 41.8 & 66.2\upred{24.4} & 60.6 & 65.9\upred{5.3} & 60.2 & 68.0\upred{7.8} & 61.3 & 64.2\upred{2.9} & 68.9 & 75.2\upred{6.3} & 62.7 & 68.3\upred{5.6} \\
\lightgraybg& Claude-3.7-Sonnet & 42.0 & 62.8\upred{20.8} & 61.6 & 67.8\upred{6.2} & 60.9 & 70.5\upred{9.6} & 64.2 & 66.1\upred{1.9} & 66.1 & 79.3\upred{13.2} & 63.2 & 70.9\upred{7.7} \\
& Claude-4-Sonnet & 61.2 & \underline{67.1}\upred{5.9} & 67.3 & \underline{69.8}\upred{2.4} & 69.2 & \textbf{72.9}\upred{3.7} & 72.9 & 74.2\upred{1.3} & 78.6 & 80.6 \upred{2.0} & 72.0 & \textbf{74.4}\upred{2.4} \\
\lightgraybg& GPT-4o-1120 & 36.2 & 55.9\upred{19.7} & 59.5 & 62.4\upred{2.9} & 55.9 & 62.3\upred{6.4} & 65.2 & 68.1\upred{2.9} & 58.6 & 75.4\upred{16.8} & 59.8 & 67.0\upred{7.2} \\
& GPT-4o-0806 & 39.4 & 58.4\upred{19.0} & 47.9 & 59.7\upred{11.8} & 46.6 & 57.0\upred{10.4} & 64.2 & 69.4\upred{5.2} & 54.9 & 67.3\upred{12.4} & 53.4 & 63.4\upred{10.0} \\
\lightgraybg& GPT-4o-mini & 28.0 & 40.8\upred{12.8} & 25.2 & 32.4\upred{7.2} & 26.8 & 32.4\upred{5.6} & 23.9 & 44.8\upred{20.9} & 9.5 & 43.4\upred{33.9} & 21.3 & 38.2\upred{16.9} \\
& GPT-4.1 & 41.4 & 59.6\upred{18.2} & 51.2 & 66.6\upred{15.4} & 53.1 & 63.5\upred{10.4} & 68.1 & 71.3\upred{3.2} & 59.8 & 68.6\upred{8.8} & 58.0 & 67.5\upred{9.5} \\
\lightgraybg& Qwen3-Coder-480B-A22B & 52.1 & \textbf{79.1}\upred{27.0} & 63.4 & 66.3\upred{2.9} & 65.5 & 69.6\upred{4.1} & \underline{74.8} & \textbf{78.1}\upred{3.2} & 74.6 & \underline{81.2}\upred{6.6} & 69.6 & 73.8\upred{4.2} \\
& Qwen2.5-Coder-32B-Instruct & 43.6 & 46.0\upred{2.4} & 36.8 & 44.9\upred{8.1} & 37.8 & 48.6\upred{10.8} & 48.4 & 68.4\upred{20.0} & 31.1 & 55.8\upred{24.7} & 38.5 & 54.4\upred{15.9} \\
\graybg& \textbf{SRI-Coder-32B} & 24.5 & 61.6\upred{37.1} & 10.3 & 60.5\upred{50.1} & 15.3 & 64.5\upred{49.2} & 5.3 & 74.9\upred{69.6} & 8.0 & 75.6\upred{67.6} & 9.8 & 69.9\upred{60.1} \\
\midrule
& \multicolumn{13}{c}{\textbf{Reasoning Chat Models}}\\
\midrule
\lightgraybg& DeepSeek-R1-0528 & 42.1 & 60.1\upred{18.0} & 62.3 & \underline{69.8}\upred{7.5} & 60.5 & \textbf{72.3}\upred{11.8} & 57.1 & \textbf{75.9}\upred{18.8} & 66.2 & \textbf{79.8}\upred{13.6} & 61.5 & \textbf{74.5}\upred{13.0} \\
& Claude-3.7-Sonnet-think & 44.2 & \underline{61.3}\upred{17.1} & 64.3 & 68.3\upred{4.0} & 63.6 & {70.7}\upred{7.1} & 58.4 & 66.8\upred{8.4} & 72.9 & \underline{79.2}\upred{6.3} & 64.8 & {71.3}\upred{6.5} \\
\lightgraybg& Grok-3 & 39.1 & \textbf{63.4}\upred{24.3} & 58.7 & 63.3\upred{4.6} & 59.1 & 66.6\upred{7.5} & 61.9 & \underline{72.3}\upred{10.4} & 65.5 & 73.8\upred{8.3} & 61.3 & 69.0\upred{7.7} \\
& o1-2024-12-17 & 41.1 & 51.5\upred{10.4} & 62.6 & 68.7\upred{6.1} & 65.9 & 71.1\upred{5.2} & 67.1 & 72.2\upred{5.1} & 65.9 & {74.4}\upred{8.5} & 65.4 & \underline{71.6}\upred{6.2} \\
\lightgraybg& o3-mini & 43.0 & 57.0\upred{14.0} & 32.1 & \textbf{76.4}\upred{44.3} & 28.0 & \underline{71.6}\upred{43.6} & 44.8 & 57.7\upred{12.9} & 38.2 & 73.1\upred{34.9} & 35.8 & 69.7\upred{34.1} \\
& Qwen3-30B-A3B & 33.0 & 48.5\upred{15.5} & 29.9 & 42.6\upred{12.7} & 34.7 & 48.0\upred{13.3} & 50.0 & 64.8\upred{14.8} & 34.1 & 58.3\upred{24.2 } & 37.2 & 53.4\upred{16.2} \\
\lightgraybg& Qwen3-235B-A22B & 35.6  & 54.7 \upred{19.1} & 33.3 & 47.5 \upred{12.6} & 36.1 & 48.7\upred{12.6} & 35.2 & 63.5\upred{28.4} & 52.6 & 58.1\upred{5.5} & 39.3 & 54.4\upred{15.2} \\

\bottomrule
\end{tabular}
\caption{Performance comparison on unit tests-based benchmarks (ExecRepoBench and SAFIM). In SAFIM, all metrics are reported as Pass@1 except API which uses Exact Match (EM).}
\vspace{-1.5em}
\label{tab:pass}
\end{table*}

\section{Experiments}
\subsection{Experimental Setup}
\paragraph{Training Dataset} To ensure robust evaluation, our training data underwent strict decontamination to prevent repository-level overlap with benchmarks. The dataset integrates two distinct components: the novel \textbf{SRI-200K} and a general instruction set. For the SRI component, we sourced code from \textit{The Stack v2} \cite{starcoder2} and employed tree-sitter to extract fundamental logic blocks (e.g., functions, loops) as infilling targets. These were formulated into search-and-replace queries with a maximum context length of 32k tokens. While we constructed 200k samples in total, we utilized a 20k subset for fine-tuning and created a markdown-formatted variant for ablation studies. Complementing this, we sampled 60k instruction pairs from \textit{Glaive-Code-Assistant} \cite{glaive_code}, meticulously filtering out completion-related tasks to maintain general instruction-following capabilities (Appendix \S\ref{app:data_construction}).

\paragraph{Benchmarks} To rigorously evaluate code completion, we employ five mainstream benchmarks originally designed for FIM tasks. These include three similarity-based metrics: CrossCodeEval \cite{crosscodeeval}, CrossCodeLongEval \cite{crosscodelongeval}, and RepoEval \cite{repoeval}, and two execution-based benchmarks: SAFIM \cite{safim} and ExecRepoBench \cite{execrepobench}. Spanning diverse sources like GitHub and Codeforces, this suite comprehensively covers both repository-level and in-file scenarios.

\paragraph{Training Details}
We fine-tuned Qwen2.5-Coder Base models using Megatron-LM \cite{megatron} on 16 NVIDIA A100 GPUs. Training utilized a 32k context length, global batch size of 256, and BFloat16 precision. The learning rate followed a linear decay from $5 \times 10^{-5}$ to $5 \times 10^{-6}$ over 853 steps. Crucially, to strictly isolate the impact of our method and prevent contamination from prior alignment, we trained starting from base checkpoints using our 80k mixed dataset (20k SRI, 60k general instructions, 100 safety). Other architectural configurations align with the standard Qwen2.5-Coder implementation \cite{qwen25coder}.

\paragraph{Evaluation Details}
We evaluated a diverse array of models, including the GPT \cite{gpt4}, Claude \cite{claude2}, Gemini \cite{gemini1.5}, DeepSeek (V3/R1) \cite{deepseekv3,deepseekr1}, and Qwen (2.5/3) \cite{qwen25,qwen3} series. Open-source models were deployed locally via vLLM (using 2x A100s for Qwen2.5-Coder), while proprietary models were accessed via official APIs. Inference utilized greedy decoding with a 256-token limit. We employed three distinct prompting strategies: standard token-based FIM for Base models, and both Natural Language prompting and our proposed SRI format for Chat models (see Figure \ref{fig:sri_prompt}, \ref{fig:prompt_standard}, \ref{fig:prompt_dialogue}, \ref{fig:prompt_template}). Context lengths were standardized to 32k for ExecRepoBench, 8k for similarity benchmarks, and full context for SAFIM. To ensure strict fairness, we maintained identical prefix/suffix contexts across all methods. Furthermore, we inserted a \roundbox{\texttt{/* MIDDLE CODE TO COMPLETE */}} identifier in SRI prompts to explicitly mark the generation target, equalizing the positional information provided implicitly by FIM and NL prompts. Further details are in Appendix \S\ref{app:detail_result}.

\begin{figure*}
    \centering
    \includegraphics[width=0.98\linewidth]{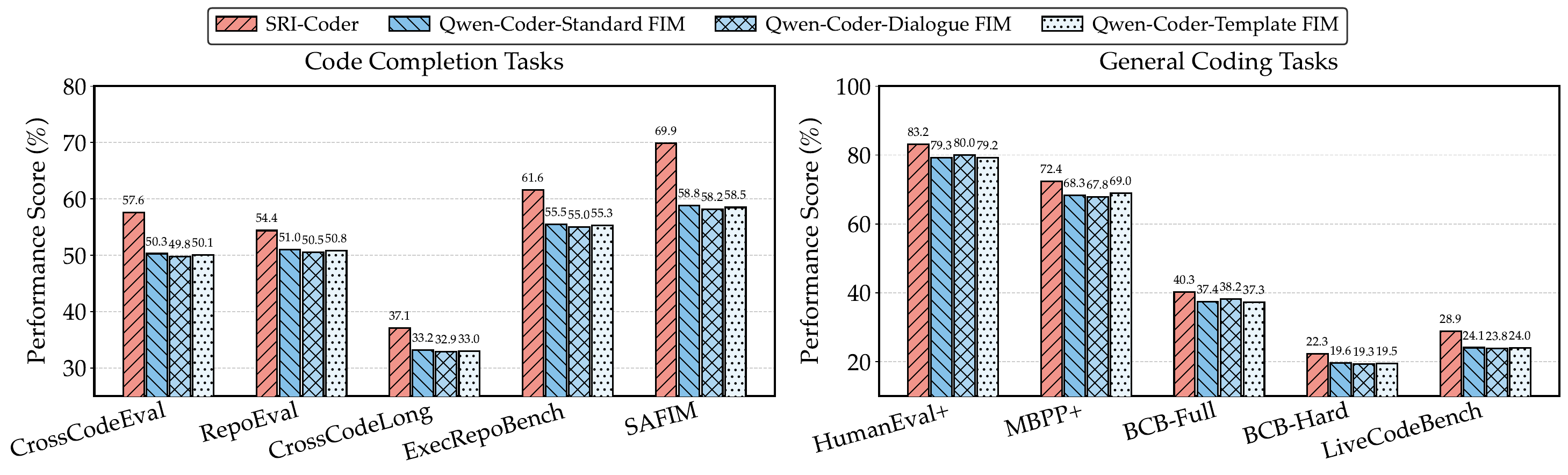}
    \caption{Ablation study comparing \textbf{SRI-Coder-32B} against three natural language-based FIM (NL-FIM) fine-tuning variants. \textbf{(Left)} On code completion benchmarks, SRI significantly outperforms the NL-FIM strategies. \textbf{(Right)} On general coding benchmarks, SRI preserves the model's broader competencies, whereas NL-FIM tuning induces measurable performance regression.}
    \vspace{-1em}
    \label{fig:32B_rank}
\end{figure*}

\subsection{Results and Analysis}
\subsubsection{Main Results}
\label{sec:main_result}
We conduct a comprehensive evaluation across five mainstream benchmarks using three distinct inference paradigms: \ding{192} Special token-based FIM (\textbf{FIM}); \ding{193} Natural language-based completion (\textbf{Chat-FIM}), for which we report the \textbf{maximum score} achieved across the three prompting variants (Standard, Dialogue, Template); and \ding{194} our proposed Search-and-Replace Infilling (\textbf{SRI}). Tables \ref{tab:ES} and \ref{tab:pass} summarize the performance across 20+ leading models (full details in Appendix \S\ref{app-sec:detail_result}).
Our analysis reveals several key findings:
\textbf{\ding{182} SRI Outperforms Chat-FIM}. SRI consistently yields significant gains over the best-performing Chat-FIM baselines. However, in a zero-shot setting, even the top Chat models still lag behind state-of-the-art (SOTA) Base models like DeepSeek-V3.
\textbf{\ding{183} Effectiveness of SRI Fine-Tuning}. Our fine-tuned \textbf{SRI-Coder-32B} surpasses its Base counterpart (Qwen2.5-Coder-32B) on all five benchmarks. Crucially, it substantially outperforms the original Instruct version across all metrics, rivaling larger models like DeepSeek-V3 and closed-source commercial APIs. This validates the high efficacy of our synthesized training data.
\textbf{\ding{184} Model Specialization}. We observe a trade-off in model specialization: reasoning-heavy models excel at short-context algorithmic tasks but often struggle with long-context, repository-level benchmarks.

\section{Discussion}
\subsection{Ablation Studies}
\label{sec:fim_sri_compare}
To isolate the efficacy of the SRI format, we conducted a rigorous comparative analysis against NL-FIM based fine-tuning strategies. We trained our primary model, \textbf{SRI-Coder-32B}, and compared it against three variants fine-tuned on the \textit{identical} 20k dataset but reformatted into distinct NL-FIM paradigms:
\ding{182} \textbf{Standard FIM Completion}: Direct instructions explicitly asking the model to complete the missing code;
\ding{183} \textbf{Dialogue-Based FIM Reconstruction}: Multi-turn interactions simulating a user seeking code restoration;
\ding{184} \textbf{Template-Based FIM Infilling}: Fixed prompt templates guiding the model to fill in specific slots.

\vspace{-0.25em}

We evaluated these models on our five code completion benchmarks alongside a suite of general coding benchmarks (MBPP, HumanEval, BigCodeBench, and LiveCodeBench). The results (Figure \ref{fig:base_chat_compare}) reveal two critical findings. First, \textbf{SRI-Coder-32B} achieves a superior average completion score of 56.1, consistently outperforming all NL-FIM variants. Second, NL-based FIM strategies result in a measurable performance regression on general coding benchmarks, whereas \textbf{SRI-Coder} effectively preserves the model's broader competencies. We hypothesize that this degradation arises because NL-FIM conditions the model to generate fragmented code segments, a pattern detrimental to the structural coherence required for general programming tasks. Consequently, we attribute SRI's superiority to its alignment with the extensive diff-formatted data inherent in pre-training corpora, which minimizes the distributional shift observed in NL-FIM strategies (analysis in Appendix \S\ref{app-sec:sri_reason_analysis}).

\vspace{-0.25em}

\begin{figure}
    \centering
    \includegraphics[width=0.98\columnwidth]{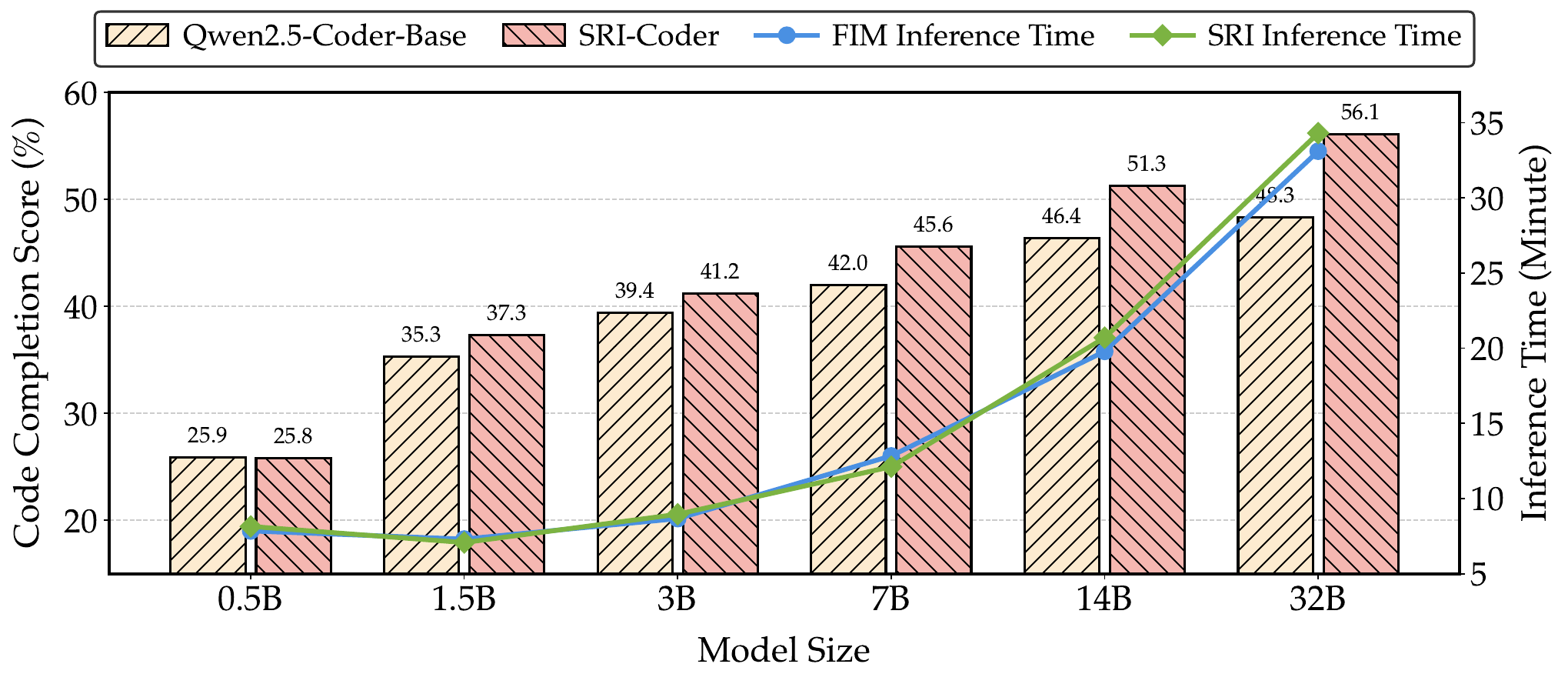}
    \caption{Scaling laws for FIM vs. SRI. Performance (left, \%) and average inference time (right, minutes) across model sizes on code completion benchmarks.}
    \vspace{-1.5em}
    \label{fig:scale}
\end{figure}

\subsection{Scaling Studies}
\label{sec:scale_study}
To investigate the effect of model scale on the performance of our SRI fine-tuning, we conducted a series of scaling experiments. Using the same mixed SFT dataset (20K SRI and 60K Glaive-Code samples), we fine-tuned the entire Qwen2.5-Coder model series, spanning parameters from 0.5B, 1.5B, 3B, 7B, 14B, to 32B.
We then evaluated two sets of models on our five code completion benchmarks: (Appendix \S\ref{app-sec:detail_result}, Table \ref{tab:sri_performance_1},\ref{tab:sri_performance_2})
\ding{182} The Qwen2.5-Coder base models, using the standard FIM inference method.
\ding{183} Our fine-tuned SRI-Coder models, using our proposed SRI inference method.
The average scores and inference times across all five benchmarks are presented in Figure \ref{fig:scale}. Our analysis yields two primary conclusions:
\textbf{\ding{182} Scaling Advantage of SRI}: The performance advantage of SRI over FIM grows as the model size increases. Larger models appear to better master the structured search-and-replace task, whereas smaller models struggle to a greater extent.
\textbf{\ding{183} Comparable Inference Speed}: For models of the same size, the inference speed of SRI is nearly identical to that of FIM.

\section{Related Works}
\subsection{Large Language Models for Coding} LLMs have demonstrated unprecedented advances in coding capabilities in recent years. Leading proprietary LLMs, such as GPT \cite{gpt4} and Claude \cite{claude}, exhibit exceptional performance across diverse programming tasks. Specialized code-centric models \cite{codex,bloom,AlphaCode,santacoder,incoder,aixcoder,codegen2,magicoder,codegemma,starcoder2,plotcraft, megaflow},, including CodeLlama \cite{code_llama}, DeepSeek-Coder \cite{deepseek_coder}, OpenCoder \cite{opencoder}, and Qwen-Coder \cite{qwen25coder}, have been developed to excel in specific tasks such as code debugging \cite{codedebug}, translation \cite{codetranslate}, and completion \cite{fim}. These models employ domain-specific architectures and are trained on extensive code corpora, comprising billions of snippets, to optimize programming-related capabilities.

\subsection{Code Completion and Coding Tools}
In the context of code completion, the pioneering work of \citet{fim} revolutionized the field by introducing the fill-in-the-middle (FIM) pre-training strategy. This approach implements a data transformation by relocating text spans from the middle of a document to its end. Models such as CodeX \cite{codex}, StarCoder \cite{starcoder2}, DeepSeek-Coder \cite{deepseek_coder}, Qwen-Coder \cite{team2025qwen3}, and CodeLlama \cite{code_llama} have integrated the FIM strategy into their training. Following this breakthrough, subsequent research has continuously advanced code completion capabilities \cite{codex, santacoder,codegen2, aixcoder} and enhanced training methodologies \cite{horizon, code_llama}. Code completion tasks require models to generate missing code segments by leveraging both left and right contexts, enabling the implementation of "\textit{autocomplete}" features in LLM-powered coding tools.

Prominent LLM-powered programming tools such as \textit{GitHub Copilot}, \textit{Augment}, and \textit{Continue.dev} have adopted FIM as their primary code completion solution, powering convenient "Tab-to-complete" functionality for developers. However, the recent rise of agentic coding has ushered in a new wave of agent-based tools, including \textit{Curosr}, \textit{Cline}, \textit{Windsurf}, \textit{Claude Code}, and \textit{Gemini-Cli}. These tools can operate at the repository level, plan implementation strategies, and perform multi-file edits. Those capabilities are far exceed the scope of simple autocomplete. This evolution necessitates a more powerful and flexible code completion paradigm than FIM, one capable of supporting more generalized and intelligent programming assistance.
A comprehensive review of additional related works is provided in Appendix \ref{sec:ad-related works}.
\section{Conclusion and Future Directions}

In this work, we identified the inherent limitations of FIM—rigidity, security vulnerabilities, and incompatibility with Chat models—and proposed \textbf{Search-and-Replace Infilling (SRI)} as a superior alternative. Through the release of our synthesized dataset and the \textbf{SRI-Coder} series, we demonstrate that minimal fine-tuning allows Chat models to surpass the completion capabilities of Base models. These findings challenge the dominance of FIM, suggesting that SRI provides a more secure and flexible foundation for next-generation, agentic coding tools.
Looking forward, we aim to validate our approach by integrating SRI-Coder into functional IDEs to gather real-world developer feedback. Additionally, observing that smaller models (e.g., 0.5B) struggle with the structural complexity of the SRI task compared to their larger counterparts, we plan to explore techniques such as knowledge distillation to effectively transfer SRI capabilities to these compact models.

\vspace{-0.5em}

\section{Limitations}
Despite the promising results, our work has two primary limitations. First, our current evaluation relies exclusively on offline benchmarks. While standardized metrics provide valuable insights, they cannot fully capture the dynamic and interactive nature of real-world development. We have not yet incorporated qualitative feedback from actual users or gathered telemetry data from functional IDE deployments. Second, we observe a performance disparity across model scales. While our SRI fine-tuning yields significant gains for larger models (e.g., 14B and 32B), the improvements are less pronounced for smaller architectures (e.g., 0.5B). This suggests that the structural complexity of the SRI task may require a certain threshold of model capacity to be fully effective.

\bibliography{custom}
\clearpage
\appendix

\renewcommand*\footnoterule{} 

\newcommand\blfootnote[1]{%
  \begingroup
  \renewcommand\thefootnote{}\footnote{#1}%
  \addtocounter{footnote}{-1}%
  \endgroup
}

\onecolumn 
\section*{Appendix}
\startcontents[sections]
\printcontents[sections]{l}{1}{\setcounter{tocdepth}{2}}
\twocolumn 

\section{Additional Related Works}
\label{sec:ad-related works}
\subsection{LLM Jailbreak}
Jailbreaking techniques for large language models (LLMs) are broadly classified into white-box and black-box methods, distinguished by the attacker’s level of access to model parameters \cite{yi2024jailbreak, ma2025safety}. White-box approaches—such as gradient-based optimization \cite{zou2024universal}, logits-driven methods \cite{guo2024cold}, and fine-tuning techniques—demonstrate high success rates but require full access to the model's internal states and parameters, limiting their practical applicability. In contrast, black-box methods operate without such access and typically rely on iterative querying. These attacks employ techniques such as template manipulation (e.g., scenario nesting \cite{ding2023wolf}, context poisoning \cite{wei2023jailbreak}, and code injection \cite{kang2024exploiting}) and prompt transformation (e.g., cryptographic encoding \cite{deng2023multilingual}, multilingual obfuscation \cite{yong2023low}, and evolutionary algorithms \cite{liu2023autodan, fakenews}). Among these diverse strategies, prompt injection remains the predominant method for black-box attacks.

Black-box prompt injection can be further categorized into deterministic one-to-one attacks \cite{yuan2023gpt, handa2024jailbreaking} and probabilistic one-to-many attacks \cite{zeng2024johnny}. While early LLMs were susceptible to simple, rule-based injections, modern models with advanced reasoning capabilities have become more resilient to such straightforward attacks. To circumvent these defenses, researchers have developed multi-step reasoning chains that systematically exploit logical vulnerabilities in LLMs, enabling sophisticated attacks against even the most advanced systems \cite{yao2025mousetrap}. This vector of attack is particularly relevant in the domain of code completion, where models utilizing the FIM paradigm are highly susceptible to exploitation through malicious prefix or suffix code injection \cite{cheng2025security}.

\subsection{Code Agents}
Recent research highlights the pivotal role of large language models (LLMs) in the development of AI agents, showcasing their capability in facilitating complex task execution through tool utilization~\cite{toolformer,babyagi,metagpt,zheng-balancing,zheng2025lagcl4rec,zheng2025negative}. Notable examples include ToolFormer, which enables tools to be used more effectively by LLMs; Meta-GPT and BabyAGI, which demonstrate advancements in autonomous task management. Studies on self-edit and self-debug have further illustrated the capacity of code models to engage in multi-round interactions for code correction and improvement.
Contemporary work also underscores the efficacy of agent systems like OpenDevin~\cite{wang2024opendevin} and SWE-Agent~\cite{swe_agent} in handling complex programming tasks at the repository level, such as SWE-Bench~\cite{swebench,swe_gym}. Additionally, \citet{agentless} introduced a lightweight agentless approach that achieves competitive performance in code editing tasks while maintaining architectural simplicity.
Notably, these agent systems commonly adopt standard diff formats for communication with LLMs, demonstrating the format's effectiveness as a universal interface for code modification tasks.

\subsection{Code Editing}
Automated code editing has progressed from early sequence-to-sequence models that translated code as flat text\cite{SequenceR,CodeEditor}, to more sophisticated structure-aware methods. These models operate on Abstract Syntax Trees (ASTs) or graphs to perform precise, local transformations \cite{CODIT}, but often ignore the broader context of a change.
More recent work directly incorporates this context. \citet{c-3po} conditions its predictions on surrounding edits, while \citet{CoEdPilot} scales this concept to the repository level by locating and propagating changes across multiple files. This principle is particularly effective for large language models (LLMs), where providing "associated edits" as few-shot examples in prompts significantly improves performance without fine-tuning \cite{grace}. Concurrently, systems like \citet{EditLord} have focused on interpretability by learning and applying explicit edit rules derived from examples.
However, many of these approaches were designed for specialized, non-LLM architectures and focus heavily on sub-tasks like edit localization. This makes them difficult to apply directly to the end-to-end inference process of modern chat-based LLMs.
\section{Detailed Results}
\label{app-sec:detail_result}
In this section, we provide a comprehensive breakdown of our experimental results, including model specifications, detailed error analyses, and granular performance metrics across varying model scales.

\subsection{Error Analysis: Gemini Series}
A primary observation during our evaluation is the significant performance degradation of the Gemini model series (e.g., Gemini-1.5-Pro-Flash, Gemini-2.0-Flash) on code completion benchmarks when applying the SRI method. While these models possess strong general reasoning capabilities, they exhibit a specific failure mode in strict instruction following for formatting.

Further analysis reveals that the Gemini series struggles to adhere to the structural constraints of the SRI format. Specifically, the models frequently fail to generate the `SEARCH` block verbatim or omit the required \roundbox{\texttt{/* MIDDLE CODE TO COMPLETE */}} identifier within the search block. We hypothesize that this behavior stems from the model overfitting to specific data formats (such as standard chat or function calling) during its post-training alignment phase, making it resistant to novel structural prompt engineering. Figure \ref{lst:gemini_error} presents a concrete example of this failure mode, where Gemini-1.5-Pro-Flash produces a search-and-replace block that entirely omits the crucial location identifier, rendering the replacement ambiguous.

\subsection{Extended Performance Metrics}
Table \ref{tab:model_params} lists the specific API versions and HuggingFace model codes used in our experiments. Tables \ref{tab:ES} and \ref{tab:pass} provide the complete performance comparisons on similarity-based and execution-based benchmarks, respectively. Additionally, to investigate the impact of model scaling on SRI effectiveness, Tables \ref{tab:sri_performance_1} and \ref{tab:sri_performance_2} detail the performance of the Qwen2.5-Coder series ranging from 0.5B to 32B parameters.

\begin{table*}[htbp] 
  \centering 
  \caption{Model code/API of the evaluated models} 

  \begin{tabular}{lrl} 
    \toprule 
    \textbf{Model} & \textbf{Params} & \textbf{Model code/API} \\ 
    \midrule 

    DeepSeek-V3-Base & 671B-A37B & \texttt{deepseek-ai/DeepSeek-V3-Base} \\
    DeepSeek-V3 & 671B-A37B & \texttt{deepseek-ai/DeepSeek-V3} \\
    DeepSeek-V3-0324 & 671B-A37B & \texttt{deepseek-ai/DeepSeek-V3-0324} \\
    DeepSeek-R1-0528 & 671B-A37B & \texttt{deepseek-ai/DeepSeek-R1-0528} \\
    \addlinespace
    Claude-3.5-Haiku & - & \texttt{claude-3-5-haiku-20241022} \\
    Claude-3.5-Sonnet & - & \texttt{claude-3-5-sonnet-20241022} \\
    Claude-3.7-Sonnet & - & \texttt{claude-3-7-sonnet-20250219} \\
    Claude-3.7-Sonnet-think & - & \texttt{claude-3-7-sonnet-20250219-thinking} \\
    Claude-4-Sonnet & - & \texttt{claude-4-sonnet-20250514} \\
    \addlinespace
    GPT-4o-1120  & - & \texttt{gpt-4o-2024-11-20} \\
    GPT-4o-0806  & - & \texttt{gpt-4o-2024-08-06} \\
    GPT-4o-mini  & - & \texttt{gpt-4o-mini-2024-08-17} \\
    GPT-4.1  & - & \texttt{gpt-4.1-2025-04-14} \\
    o1  & - & \texttt{o1-2024-12-17} \\
    o3-mini  & - & \texttt{o3-mini-2025-01-31} \\
    \addlinespace 
    Gemini-1.5-Pro-Flash & - & \texttt{gemini-1.5-pro-flash} \\
    Gemini-1.5-Pro & - & \texttt{gemini-1.5-pro} \\
    Gemini-2.0-Flash & - & \texttt{gemini-2.0-flash} \\
    Gemini-2.0-Flash-Think & - & \texttt{gemini-2.0-flash-thinking} \\
    \addlinespace
    Grok-3 & - & \texttt{grok-3} \\
    \addlinespace
    Qwen2.5-Coder-32B & 32B & \texttt{Qwen/Qwen2.5-Coder-32B} \\
    Qwen2.5-Coder-32B-Instruct & 32B & \texttt{Qwen/Qwen2.5-Coder-32B-Instruct} \\
    Qwen3-30B-A3B & 30B-A3B & \texttt{Qwen/Qwen3-30B-A3B} \\
    Qwen3-235B-A22B & 235B-A22B & \texttt{Qwen/Qwen3-235B-A22B} \\
    Qwen3-Coder-480B-A35B & 480B-A35B & \texttt{Qwen/Qwen3-Coder-480B-A35B-Instruct} \\

    \bottomrule 
  \end{tabular}
  \label{tab:model_params}
\end{table*}

\begin{figure*}
\begin{tcolorbox}[title={Error search replace case for Gemini}]
\begin{lstlisting}[breaklines=true, basicstyle=\ttfamily]
    
```replace
<<<<<<< SEARCH
    """
    Preprocess a mask for the model.
    """
    mask = mask.convert("L")
    w, h = mask.size
    w, h = map(lambda x: x - x \% 32, (w, h))  # resize to integer multiple of 32
    mask = mask.resize(
        (w // scale_factor, h // scale_factor), resample=PIL.Image.NEAREST
    )
    mask = np.array(mask).astype(np.float32) / 255.0
    mask = np.tile(mask, (4, 1, 1))
    mask = mask[None].transpose(0, 1, 2, 3)  # what does this step do?
    mask = 1 - mask  # repaint white, keep black
    mask = torch.from_numpy(mask)
    return mask
=======
    """
    Preprocess a mask for the model.
    """
    mask = mask.convert("L")
    w, h = mask.size
    w, h = map(lambda x: x - x % 32, (w, h))  # resize to integer multiple of 32
    mask = mask.resize(
        (w // scale_factor, h // scale_factor), resample=PIL.Image.NEAREST
    )
    mask = np.array(mask).astype(np.float32) / 255.0
    mask = np.tile(mask, (4, 1, 1))
    mask = mask[None].transpose(0, 1, 2, 3)  # what does this step do?
    mask = 1 - mask  # repaint white, keep black
    mask = torch.from_numpy(mask)
    return mask

>>>>>>> REPLACE
```
\end{lstlisting}

\end{tcolorbox}
\caption{Error search replace case for Gemini}
\label{lst:gemini_error}
\end{figure*}

\begin{table*}[h]
\centering
\small
\setlength{\tabcolsep}{2pt}
\begin{tabular}{@{}llcccccccccccc@{}}
\toprule
& \multirow{3}{*}{\textbf{Model}} & \multicolumn{4}{c}{\textbf{CrossCodeEval}} & \multicolumn{4}{c}{\textbf{RepoEval}} & \multicolumn{4}{c}{\textbf{CrossCodeLongEval}} \\
\cmidrule(lr){3-6} \cmidrule(lr){7-10} \cmidrule(lr){11-14}
& & \multicolumn{2}{c}{\textit{EM}} & \multicolumn{2}{c}{\textit{ES}} & \multicolumn{2}{c}{\textit{EM}} & \multicolumn{2}{c}{\textit{ES}} & \multicolumn{2}{c}{\textit{EM}} & \multicolumn{2}{c}{\textit{ES}} \\
\cmidrule(lr){3-4} \cmidrule(lr){5-6} \cmidrule(lr){7-8} \cmidrule(lr){9-10} \cmidrule(lr){11-12} \cmidrule(lr){13-14}
& & \textbf{FIM} & \textbf{SRI} & \textbf{FIM} & \textbf{SRI} & \textbf{FIM} & \textbf{SRI} & \textbf{FIM} & \textbf{SRI} & \textbf{FIM} & \textbf{SRI} & \textbf{FIM} & \textbf{SRI} \\
\midrule
& \multicolumn{13}{c}{\textbf{Base Models}}\\
\midrule
\lightgraybg& Qwen2.5-Coder-32B & 57.1 & - & 86.8 & - & 51.6 & - & 78.5 & - & 36.8 & - & 66.4 & - \\
& DeepSeek-V3-Base & 61.9 & - & 88.5 & - & 63.6 & - & 86.2 & - & 50.5 & - & 77.9 & - \\
\midrule
& \multicolumn{13}{c}{\textbf{Non-reasoning Chat Models}}\\
\midrule
\lightgraybg& DeepSeek-V3 & 39.0 & 56.1\upred{17.1} & 70.3 & 86.2\upred{15.9} & 37.7 & 54.3\upred{16.6} & 60.4 & 78.9\upred{18.5} & 25.4 & 34.8\upred{9.4} & 53.5 & 67.4\upred{13.9} \\
& DeepSeek-V3-0324 & 44.5 & 56.0\upred{11.5} & 80.0 & 85.7\upred{5.7} & 45.8 & 54.5\upred{8.7} & 68.2 & {79.3}\upred{11.1} & 30.9 & 34.8\upred{3.9} & 60.2 & 68.2\upred{8.0} \\
\lightgraybg& Claude-3.5-Haiku & 23.9 & 44.5\upred{20.6} & 73.2 & 75.2\upred{2.0} & 31.9 & 44.4\upred{12.5} & 58.2 & 67.8\upred{9.6} & 17.7 & 27.1\upred{9.4} & 49.8 & 56.2\upred{6.4} \\
& Claude-3.5-Sonnet & 47.5 & 56.0\upred{8.5} & 82.3 & 85.6\upred{3.3} & 48.0 & 54.4\upred{6.4} & 72.1 & 80.2\upred{8.1} & 32.4 & 36.3\upred{3.9} & 63.2 & 69.4\upred{6.2} \\
\lightgraybg& Claude-3.7-Sonnet & 48.3 & 56.7\upred{8.4} & 81.6 & 85.7\upred{4.1} & 49.3 & 55.0\upred{5.7} & 73.6 & 80.2\upred{6.6} & 32.5 & 38.2\upred{5.7} & 64.0 & \underline{70.5}\upred{6.5} \\
& Claude-4-Sonnet & 57.5 & \underline{60.8}\upred{3.3} & \underline{87.3} & 87.2\downblue{0.1} & 55.7 & \underline{58.9}\upred{3.2} & 78.7 & \underline{82.0}\upred{3.3} & 38.1 & \underline{38.9}\upred{0.7} & 68.8 & 69.5\upred{0.7} \\
\lightgraybg& GPT-4o-1120 & 34.2 & 44.1\upred{9.9} & 72.5 & 80.9\upred{8.4} & 25.7 & 45.0\upred{19.3} & 52.1 & 72.6\upred{20.5} & 21.7 & 23.6\upred{1.9} & 51.6 & 56.4\upred{4.8} \\
& GPT-4o-0806 & 40.3 & 41.7\upred{1.4} & 77.3 & 79.5\upred{2.2} & 37.2 & 46.8\upred{9.6} & 61.5 & 73.7\upred{12.2} & 20.6 & 26.4\upred{5.8} & 50.7 & 58.2\upred{7.5} \\

\lightgraybg& GPT-4o-mini & 9.0 & 27.7\upred{18.7} & 59.6 & 68.3\upred{8.7} & 19.6 & 27.7\upred{8.1} & 48.0 & 54.6\upred{6.6} & 12.1 & 11.1\downblue{1.0} & 41.3 & 39.6\downblue{1.7} \\
& GPT-4.1 & 46.1 & 45.8\downblue{0.3} & 81.0 & 79.8\downblue{1.2} & 47.0 & 46.9\downblue{0.1} & 71.7 & 73.2\upred{1.5} & 30.8 & 28.5\downblue{2.3} & 62.2 & 61.2\downblue{1.0} \\
\lightgraybg& Gemini-1.5-Pro-Flash & 42.7 & 30.9\downblue{11.8} & 80.2 & 58.7\downblue{21.5} & 42.3 & 37.6\downblue{4.7} & 70.6 & 60.3\downblue{10.3} & 26.0 & 9.8\downblue{16.2} & 57.5 & 27.9\downblue{29.6} \\
& Gemini-1.5-Pro & 49.6 & 40.8\downblue{8.8} & 83.3 & 68.6\downblue{14.7} & 50.4 & 44.1\downblue{6.3} & 75.8 & 69.3\downblue{6.5} & 33.7  & 28.3\downblue{5.4} & 64.4 & 57.6\downblue{6.8} \\
\lightgraybg& Gemini-2.0-Flash & 43.8 & 6.8\downblue{37.0} & 75.7 & 16.5\downblue{59.2} & 44.4 & 9.0\downblue{35.4} & 67.2 & 15.1\downblue{52.1} & 29.0 & 4.4\downblue{24.6} & 56.8 & 14.3\downblue{42.5} \\

& Qwen3-Coder-480B-A35B & 56.1 & \textbf{62.9}\upred{6.9}  & 85.8 & \textbf{89.5}\upred{3.7} & 51.1 & \textbf{65.1}\upred{14.0} & 73.2 & \textbf{85.4}\upred{12.2} & 36.2 & \textbf{45.9} \upred{9.7} & 65.3 & \textbf{74.7}\upred{9.4} \\
\lightgraybg& Qwen2.5-Coder-32B-Instruct & 23.4 & 44.4\upred{21.0} & 69.9 & 76.8\upred{6.9} & 31.1 & 44.8\upred{13.7} & 56.1 & 69.7\upred{13.6} & 20.4 & 25.9\upred{5.5} & 50.3 & 55.8\upred{5.5} \\
\graybg& \textbf{SRI-Coder-32B} & 11.3 & 57.6\upred{46.3} & 37.9 & \underline{87.3}\upred{49.3} & 17.3 & 54.4\upred{37.1} & 37.3 & 78.9\upred{41.5} & 14.6 & 37.1\upred{22.5} & 33.2 & 67.9\upred{34.7} \\
\midrule
& \multicolumn{13}{c}{\textbf{Reasoning Chat Models}}\\
\midrule
\lightgraybg& DeepSeek-R1-0528 & 44.6 & 51.1\upred{6.5} & 82.3 & 84.0\upred{1.7} & 48.1 & 49.2\upred{1.1} & 75.1 & 76.0\upred{0.9} & 29.2 & 29.7\upred{0.5} & 62.2 & 62.6\upred{0.4} \\
& Claude-3.7-Sonnet-think & 49.4 & \textbf{54.6}\upred{5.2} & 82.5 & \textbf{84.5}\upred{2.0} & 9.1 & \underline{53.0}\upred{43.9} & 27.6 & \underline{78.0}\upred{50.4} & 30.6 & \underline{34.9}\upred{4.3} & 60.7 & \underline{66.9}\upred{6.2} \\
\lightgraybg& Gemini-2.0-Flash-think & 37.4 & 32.9\downblue{4.5} & 61.3 & 62.3\upred{1.0} & 46.3 & 44.8\downblue{1.5} & 68.9 & 64.9\downblue{4.0} & 28.9 & 21.2\downblue{7.7} & 56.4 & 47.4\downblue{9.0} \\
& Grok-3 & 46.4 & \underline{54.4}\upred{8.0} & 82.1 & \underline{84.1}\upred{2.0} & 50.1 & \textbf{53.6}\upred{3.5} & 75.8 & \textbf{78.7}\upred{2.9} & 32.4 & \textbf{35.2}\upred{2.8} & 65.2 & \textbf{68.0}\upred{2.8} \\
\lightgraybg& o1-2024-12-17 & 26.9 & 42.7\upred{15.8} & 75.4 & 78.4\upred{3.0} & 44.5 & 48.3\upred{3.8} & 68.4 & 73.6\upred{5.2} & 26.6 & 24.4\downblue{2.2} & 56.6 & 57.0\upred{0.4} \\
& o3-mini & 23.7 & 27.6\upred{3.9} & 39.1 & 69.9\upred{30.8} & 26.4 & 39.1\upred{12.7} & 37.1 & 61.4\upred{24.3} & 9.7 & 20.9\upred{11.2} & 21.1 & 48.9\upred{27.8} \\
\lightgraybg& Qwen3-30B-A3B & 13.7 & 36.1\upred{22.4} & 59.7 & 71.5\upred{11.8} & 20.0 & 33.6\upred{13.6} & 44.3 & 61.2\upred{16.9} & 12.9 & 17.6\upred{4.7} & 36.7 & 45.1\upred{8.4} \\
& Qwen3-235B-A22B & 14.6  & 42.7\upred{28.2} & 35.6 & 76.8\upred{41.3} & 19.1 & 42.7\upred{23.6} & 30.6 & 68.9\upred{38.3} & 15.0 & 23.2 \upred{8.2} & 32.5 & 53.5\upred{21.0} \\

\bottomrule
\end{tabular}
\caption{Performance comparison on similarity-based benchmarks (CrossCodeEval, RepoEval, and CrossCodeLongEval). For base models, \textbf{FIM} denotes special token based fill-in-the-middle code completion, for chat models, \textbf{FIM} denotes natural language-based code completion prompting, while \textbf{SRI} represents our method. \textbf{Bold} and \underline{underlined} values indicate the best and second-best performance respectively within the model type. \upred{n} and \downblue{n} indicate performance improvements and degradations of SRI compared to FIM.}
\label{tab:ES}
\vspace{-10pt}
\end{table*}

\begin{table*}[h]
\centering
\small
\setlength{\tabcolsep}{2pt}
\begin{tabular}{@{}llcccccccccccc@{}}
\toprule
& \multirow{3}{*}{\textbf{Model}} & \multicolumn{2}{c}{\textbf{ExecRepoBench}} & \multicolumn{10}{c}{\textbf{SAFIM}} \\
\cmidrule(lr){3-4} \cmidrule(lr){5-14}
& & \multicolumn{2}{c}{\textit{Pass@1}} & \multicolumn{2}{c}{\textit{Block}} & \multicolumn{2}{c}{\textit{Block V2}} & \multicolumn{2}{c}{\textit{API}} & \multicolumn{2}{c}{\textit{Control}} & \multicolumn{2}{c}{\textit{Average}} \\
\cmidrule(lr){3-4} \cmidrule(lr){5-6} \cmidrule(lr){7-8} \cmidrule(lr){9-10} \cmidrule(lr){11-12} \cmidrule(lr){13-14}
& & \textbf{FIM} & \textbf{SRI} & \textbf{FIM} & \textbf{SRI} & \textbf{FIM} & \textbf{SRI} & \textbf{FIM} & \textbf{SRI} & \textbf{FIM} & \textbf{SRI} & \textbf{FIM} & \textbf{SRI} \\
\midrule
& \multicolumn{13}{c}{\textbf{Base Models}}\\
\midrule
\lightgraybg& Qwen2.5-Coder-32B & 25.7    & - & 62.9  & - & 65.2   & - & 75.5   & - & 76.4  & - & 70.0  & - \\
& DeepSeek-V3-Base & 36.5  & - & 76.3  & - & 77.9  & - & 79.4  & - & 84.8  & - & 79.6  & - \\
\midrule
& \multicolumn{13}{c}{\textbf{Non-reasoning Chat Models}}\\
\midrule
\lightgraybg& DeepSeek-V3 & 35.6 & 61.8\upred{26.2} & 60.5 & \textbf{70.1}\upred{9.6} & 59.7 & 70.6\upred{10.9} & 55.8 & 73.6\upred{17.8} & 64.1 & 80.6\upred{16.5} & 60.0 & 73.7\upred{13.7} \\
& DeepSeek-v3-0324 & 36.9 & 65.5\upred{28.6} & 53.3 & \underline{69.8}\upred{16.5} & 56.5 & \underline{71.1}\upred{14.6} & 68.1 & 74.5\upred{6.4} & 65.3 & \textbf{81.8}\upred{16.5} & 60.8 & \underline{74.3}\upred{13.5} \\
\lightgraybg& Claude-3.5-Haiku & 35.6 & 57.7\upred{22.1} & 60.1 & 66.3\upred{6.2} & 59.7 & 68.0\upred{8.3} & 65.2 & 71.6\upred{6.4} & 68.5 & 77.4\upred{8.9} & 63.4 & 70.8\upred{7.4} \\
& Claude-3.5-Sonnet & 41.8 & 66.2\upred{24.4} & 60.6 & 65.9\upred{5.3} & 60.2 & 68.0\upred{7.8} & 61.3 & 64.2\upred{2.9} & 68.9 & 75.2\upred{6.3} & 62.7 & 68.3\upred{5.6} \\
\lightgraybg& Claude-3.7-Sonnet & 42.0 & 62.8\upred{20.8} & 61.6 & 67.8\upred{6.2} & 60.9 & 70.5\upred{9.6} & 64.2 & 66.1\upred{1.9} & 66.1 & 79.3\upred{13.2} & 63.2 & 70.9\upred{7.7} \\
& Claude-4-Sonnet & 61.2 & \underline{67.1}\upred{5.9} & 67.3 & \underline{69.8}\upred{2.4} & 69.2 & \textbf{72.9}\upred{3.7} & 72.9 & 74.2\upred{1.3} & 78.6 & 80.6 \upred{2.0} & 72.0 & \textbf{74.4}\upred{2.4} \\
\lightgraybg& GPT-4o-1120 & 36.2 & 55.9\upred{19.7} & 59.5 & 62.4\upred{2.9} & 55.9 & 62.3\upred{6.4} & 65.2 & 68.1\upred{2.9} & 58.6 & 75.4\upred{16.8} & 59.8 & 67.0\upred{7.2} \\
& GPT-4o-0806 & 39.4 & 58.4\upred{19.0} & 47.9 & 59.7\upred{11.8} & 46.6 & 57.0\upred{10.4} & 64.2 & 69.4\upred{5.2} & 54.9 & 67.3\upred{12.4} & 53.4 & 63.4\upred{10.0} \\
\lightgraybg& GPT-4o-mini & 28.0 & 40.8\upred{12.8} & 25.2 & 32.4\upred{7.2} & 26.8 & 32.4\upred{5.6} & 23.9 & 44.8\upred{20.9} & 9.5 & 43.4\upred{33.9} & 21.3 & 38.2\upred{16.9} \\
& GPT-4.1 & 41.4 & 59.6\upred{18.2} & 51.2 & 66.6\upred{15.4} & 53.1 & 63.5\upred{10.4} & 68.1 & 71.3\upred{3.2} & 59.8 & 68.6\upred{8.8} & 58.0 & 67.5\upred{9.5} \\
\lightgraybg& Gemini-1.5-Pro-Flash & 32.7 & 45.8\upred{13.1} & 50.0 & 30.5\downblue{19.5} & 49.6 & 32.3\downblue{17.3} & 65.2 & 54.5\downblue{10.7} & 51.1 & 33.9\downblue{17.2} & 54.0 & 37.8\downblue{16.2} \\
& Gemini-1.5-Pro & 37.9 & 53.4\upred{15.5} & 60.8 & 56.0\downblue{4.8} & 56.3 & 53.2\downblue{3.1} & 69.4 & 67.4\downblue{2.0} & 57.7 & 56.7\downblue{1.0} & 61.0 & 58.3\downblue{2.7} \\
\lightgraybg& Gemini-2.0-Flash & 41.1 & 31.1\downblue{10.0} & 62.2 & 19.7\downblue{42.5} & 58.0 & 17.6\downblue{40.4} & 66.8 & 24.2\downblue{42.6} & 64.1 & 17.8\downblue{46.3} & 62.8 & 19.8\downblue{43.0} \\
& Qwen3-Coder-480B-A22B & 52.1 & \textbf{79.1}\upred{27.0} & 63.4 & 66.3\upred{2.9} & 65.5 & 69.6\upred{4.1} & \underline{74.8} & \textbf{78.1}\upred{3.2} & 74.6 & \underline{81.2}\upred{6.6} & 69.6 & 73.8\upred{4.2} \\
\lightgraybg& Qwen2.5-Coder-32B-Instruct & 43.6 & 46.0\upred{2.4} & 36.8 & 44.9\upred{8.1} & 37.8 & 48.6\upred{10.8} & 48.4 & 68.4\upred{20.0} & 31.1 & 55.8\upred{24.7} & 38.5 & 54.4\upred{15.9} \\
\graybg& \textbf{SRI-Coder-32B} & 24.5 & 61.6\upred{37.1} & 10.3 & 60.5\upred{50.1} & 15.3 & 64.5\upred{49.2} & 5.3 & 74.9\upred{69.6} & 8.0 & 75.6\upred{67.6} & 9.8 & 69.9\upred{60.1} \\
\midrule
& \multicolumn{13}{c}{\textbf{Reasoning Chat Models}}\\
\midrule
\lightgraybg& DeepSeek-R1-0528 & 42.1 & 60.1\upred{18.0} & 62.3 & \underline{69.8}\upred{7.5} & 60.5 & \textbf{72.3}\upred{11.8} & 57.1 & \textbf{75.9}\upred{18.8} & 66.2 & \textbf{79.8}\upred{13.6} & 61.5 & \textbf{74.5}\upred{13.0} \\
& Claude-3.7-Sonnet-think & 44.2 & \underline{61.3}\upred{17.1} & 64.3 & 68.3\upred{4.0} & 63.6 & {70.7}\upred{7.1} & 58.4 & 66.8\upred{8.4} & 72.9 & \underline{79.2}\upred{6.3} & 64.8 & {71.3}\upred{6.5} \\
\lightgraybg& Gemini-2.0-Flash-think & 42.7 & 42.2\downblue{0.5} & 35.2 & 33.7\downblue{1.5} & 36.1 & 34.0\downblue{2.1} & 63.6 & 67.7\upred{4.1} & 43.7 & 38.5\downblue{5.2} & 44.6 & 43.5\downblue{1.1} \\
& Grok-3 & 39.1 & \textbf{63.4}\upred{24.3} & 58.7 & 63.3\upred{4.6} & 59.1 & 66.6\upred{7.5} & 61.9 & \underline{72.3}\upred{10.4} & 65.5 & 73.8\upred{8.3} & 61.3 & 69.0\upred{7.7} \\
\lightgraybg& o1-2024-12-17 & 41.1 & 51.5\upred{10.4} & 62.6 & 68.7\upred{6.1} & 65.9 & 71.1\upred{5.2} & 67.1 & 72.2\upred{5.1} & 65.9 & {74.4}\upred{8.5} & 65.4 & \underline{71.6}\upred{6.2} \\
& o3-mini & 43.0 & 57.0\upred{14.0} & 32.1 & \textbf{76.4}\upred{44.3} & 28.0 & \underline{71.6}\upred{43.6} & 44.8 & 57.7\upred{12.9} & 38.2 & 73.1\upred{34.9} & 35.8 & 69.7\upred{34.1} \\
\lightgraybg& Qwen3-30B-A3B & 33.0 & 48.5\upred{15.5} & 29.9 & 42.6\upred{12.7} & 34.7 & 48.0\upred{13.3} & 50.0 & 64.8\upred{14.8} & 34.1 & 58.3\upred{24.2 } & 37.2 & 53.4\upred{16.2} \\
& Qwen3-235B-A22B & 35.6  & 54.7 \upred{19.1} & 33.3 & 47.5 \upred{12.6} & 36.1 & 48.7\upred{12.6} & 35.2 & 63.5\upred{28.4} & 52.6 & 58.1\upred{5.5} & 39.3 & 54.4\upred{15.2} \\

\bottomrule
\end{tabular}
\caption{Performance comparison on unit tests-based benchmarks (ExecRepoBench and SAFIM). In SAFIM, all metrics are reported as Pass@1 except API which uses Exact Match (EM).}
\vspace{-7pt}
\label{tab:pass}
\end{table*}

\begin{table*}[h]
\centering
\small 
\setlength{\tabcolsep}{4pt} 

\begin{tabular}{@{}llcccccccccccc@{}}
\toprule
& \multirow{3}{*}{\textbf{Model}} & \multicolumn{4}{c}{\textbf{CrossCodeEval}} & \multicolumn{4}{c}{\textbf{RepoEval}} & \multicolumn{4}{c}{\textbf{CrossCodeLongEval}} \\
\cmidrule(lr){3-6} \cmidrule(lr){7-10} \cmidrule(lr){11-14} 

& & \multicolumn{2}{c}{\textit{EM}} & \multicolumn{2}{c}{\textit{ES}} & \multicolumn{2}{c}{\textit{EM}} & \multicolumn{2}{c}{\textit{ES}} & \multicolumn{2}{c}{\textit{EM}} & \multicolumn{2}{c}{\textit{ES}} \\
\cmidrule(lr){3-4} \cmidrule(lr){5-6} \cmidrule(lr){7-8} \cmidrule(lr){9-10} \cmidrule(lr){11-12} \cmidrule(lr){13-14} 

& & \textbf{FIM} & \textbf{SRI} & \textbf{FIM} & \textbf{SRI} & \textbf{FIM} & \textbf{SRI} & \textbf{FIM} & \textbf{SRI} & \textbf{FIM} & \textbf{SRI} & \textbf{FIM} & \textbf{SRI} \\
\midrule 

\lightgraybg & Qwen2.5-Coder-0.5B        & 24.6 & - & 68.9 & - & 28.1 & - & 63.0 & - & 19.7 & - & 51.1 & - \\
             & SRI-Coder-0.5B            & - & 24.4 & - & 66.5 & - & 25.2 & - & 57.3 & - & 16.4 & - & 48.1 \\
\lightgraybg & Qwen2.5-Coder-1.5B        & 40.2 & - & 78.2 & - & 40.5 & - & 71.7 & - & 28.3 & - & 59.2 & - \\
             & SRI-Coder-1.5B            & - & 39.7 & - & 76.6 & - & 39.4 & - & 68.8 & - & 29.2 & - & 60.5 \\
\lightgraybg & Qwen2.5-Coder-3B          & 44.9 & - & 80.9 & - & 44.0 & - & 74.0 & - & 30.1 & - & 61.3 & - \\
             & SRI-Coder-3B              & - & 43.5 & - & 79.8 & - & 44.3 & - & 74.6 & - & 30.9 & - & 61.9 \\
\lightgraybg & Qwen2.5-Coder-7B          & 49.3 & - & 83.1 & - & 46.3 & - & 75.1 & - & 33.4 & - & 63.9 & - \\
             & SRI-Coder-7B              & - & 50.3 & - & 84.0 & - & 46.5 & - & 74.0 & - & 34.5 & - & 62.2 \\
\lightgraybg & Qwen2.5-Coder-14B         & 55.4 & - & 86.1 & - & 50.6 & - & 79.0 & - & 36.2 & - & 65.8 & - \\
             & SRI-Coder-14B             & - & 53.3 & - & 85.5 & - & 52.4 & - & 78.2 & - & 34.0 & - & 66.2 \\
\lightgraybg & Qwen2.5-Coder-32B         & 57.1 & - & 86.8 & - & 51.6 & - & 78.5 & - & 36.9 & - & 66.5 & - \\
             & SRI-Coder-32B             & - & 57.6 & - & 87.1 & - & 54.4 & - & 78.9 & - & 37.1 & - & 67.9 \\

\bottomrule 
\end{tabular}
\caption{Performance of Qwen2.5-Coder models with and without SRI on various benchmarks. For each model pair, the first row shows FIM performance, and the second row shows SRI performance.}
\label{tab:sri_performance_1}
\end{table*}

\begin{table*}[h]
\centering
\small 
\setlength{\tabcolsep}{4pt} 

\begin{tabular}{@{}llcccccccccccc@{}}
\toprule
& \multirow{3}{*}{\textbf{Model}} & \multicolumn{2}{c}{\textbf{ExecRepoBench}} & \multicolumn{10}{c}{\textbf{SAFIM}} \\
\cmidrule(lr){3-4} \cmidrule(lr){5-14} 

& & \multicolumn{2}{c}{\textit{Pass@1}} & \multicolumn{2}{c}{\textit{Block}} & \multicolumn{2}{c}{\textit{Block V2}} & \multicolumn{2}{c}{\textit{API}} & \multicolumn{2}{c}{\textit{Control}} & \multicolumn{2}{c}{\textit{Average}} \\
\cmidrule(lr){3-4} \cmidrule(lr){5-6} \cmidrule(lr){7-8} \cmidrule(lr){9-10} \cmidrule(lr){11-12} \cmidrule(lr){13-14} 

& & \textbf{FIM} & \textbf{SRI} & \textbf{FIM} & \textbf{SRI} & \textbf{FIM} & \textbf{SRI} & \textbf{FIM} & \textbf{SRI} & \textbf{FIM} & \textbf{SRI} & \textbf{FIM} & \textbf{SRI} \\
\midrule 

\lightgraybg & Qwen2.5-Coder-0.5B        & 22.0 & - & 24.6 & - & 30.5 & - & 47.7 & - & 37.2 & - & 35.0 & - \\
             & SRI-Coder-0.5B            & - & 36.8 & - & 15.0 & - & 20.3 & - & 41.9 & - & 23.8 & - & 25.3 \\
\lightgraybg & Qwen2.5-Coder-1.5B        & 17.2 & - & 38.9 & - & 44.8 & - & 64.8 & - & 52.6 & - & 50.3 & - \\
             & SRI-Coder-1.5B            & - & 46.0 & - & 33.3 & - & 39.8 & - & 64.4 & - & 48.2 & - & 46.4 \\
\lightgraybg & Qwen2.5-Coder-3B          & 22.3 & - & 46.6 & - & 50.3 & - & 65.8 & - & 60.3 & - & 55.8 & - \\
             & SRI-Coder-3B              & - & 48.6 & - & 43.3 & - & 53.9 & - & 68.5 & - & 57.7 & - & 55.9 \\
\lightgraybg & Qwen2.5-Coder-7B          & 19.8 & - & 52.2 & - & 56.8 & - & 70.6 & - & 65.2 & - & 61.2 & - \\
             & SRI-Coder-7B              & - & 51.5 & - & 44.6 & - & 50.6 & - & 69.1 & - & 61.7 & - & 57.0 \\
\lightgraybg & Qwen2.5-Coder-14B         & 22.6 & - & 58.8 & - & 62.5 & - & 74.5 & - & 72.3 & - & 67.0 & - \\
             & SRI-Coder-14B             & - & 59.7 & - & 58.1 & - & 63.6 & - & 73.0 & - & 66.0 & - & 67.2 \\
\lightgraybg & Qwen2.5-Coder-32B         & 25.7 & - & 62.9 & - & 65.2 & - & 75.5 & - & 76.4 & - & 70.0 & - \\
             & SRI-Coder-32B             & - & 61.6 & - & 60.5 & - & 64.5 & - & 74.9 & - & 75.6 & - & 69.9 \\

\bottomrule 
\end{tabular}
\caption{Performance comparison of Qwen2.5-Coder models on unit test-based benchmarks (ExecRepoBench and SAFIM). For each model size, FIM and SRI performances are shown in separate rows.}
\label{tab:sri_performance_2}
\end{table*}

\section{Experiments Details}
\label{app:detail_result}

\subsection{Training Dataset}
Our training dataset consists of two main components: a novel Search-and-Replace Infilling (SRI) dataset and a conventional code instruction dataset. To ensure data integrity and prevent leakage, our training and benchmark datasets underwent a strict decontamination process, and we avoided any repository-level overlap between them.
\textbf{SRI Data}. We constructed the SRI dataset using the open-source, repository-level dataset The Stack v2 \cite{starcoder2}. To generate training samples, we first parse code files using tree-sitter to identify and extract fundamental logic blocks (e.g., functions, classes, loops). These blocks serve as the "middle" segments for our code infilling tasks. For each extracted block, we formulate a multi-level code completion query using its surrounding code as context. The ground truth is then formatted into a standard search-and-replace structure. This process yielded a large-scale dataset of 200K SRI samples, which we call SRI-200K, with a maximum context length of 32,768 tokens. From this dataset, we sampled a 20K subset to fine-tune our Qwen2.5-Coder-SRI models. For our ablation studies, we created a parallel version of this 20K subset where the ground truth was reformatted into a standard markdown code block, serving as the target output for natural language-based code completion prompts.
\textbf{Instruction Data}. The second component is a general-purpose code instruction dataset. We sampled 60K question-answer pairs from the Glaive-Code-Assistant dataset \cite{glaive_code}. To ensure a clear separation of tasks, we meticulously filtered this sample to remove all data related to code completion or search-and-replace tasks.

\subsection{Benchmarks}
To evaluate the code completion capabilities of the models, we selected five mainstream benchmarks originally designed for Fill-in-the-Middle (FIM) evaluation with base LLMs. Our selection includes three benchmarks based on similarity metrics including CrossCodeEval \cite{crosscodeeval}, CrossCodeLongEval \cite{crosscodelongeval}, and RepoEval \cite{repoeval} and two based on unit test execution: SAFIM \cite{safim} and ExecRepoBench \cite{execrepobench}.
These benchmarks were chosen for their diversity and rigor. They cover a wide range of scenarios, including both repository-level and in-file completion tasks. Furthermore, their data is sourced from diverse origins, spanning popular GitHub repositories and algorithm code from Codeforces. This comprehensive set of benchmarks allows us to thoroughly assess the models' overall code completion performance under various conditions.

\vspace{-2pt}

\subsection{Training Details}
All models were trained on 16 NVIDIA A100 (80GB) GPUs using the Megatron-LM framework\cite{megatron}. We configured the training with a context length of 32,768 tokens, a global batch size of 256, and a micro-batch size of 1. The learning rate was set to $5 \times 10^{-5}$, with a warm-up phase of 30 steps and a minimum learning rate of $5 \times 10^{-6}$, followed by a linear decay over 853 steps. We used BFloat16(BF16) for mixed-precision training.
The model architecture employs the SwiGLU \cite{SwiGLU} activation function and RMSNorm \cite{RMSNorm} (with $\epsilon = 10^{-6}$) for normalization. Gradient clipping was set to 1.0, and a weight decay of 0.1 was applied. The total training dataset comprised 80K samples: 20K for our SRI code completion task, 60K for general code-related instruction following, and 100 samples for safety alignment.
To avoid potential data contamination, we fine-tuned our models starting from the pre-trained Qwen2.5-Coder base models rather than the instruction-tuned Qwen2.5-Coder-Instruct series. This decision prevents any pre-existing exposure to code completion or search-and-replace SFT data in the instruction-tuned models from confounding our experimental results. For further details on the model architecture, please refer to \citet{qwen25coder}.

\subsection{Evaluation Details}
Our evaluation encompasses a diverse set of models, including the GPT series \cite{gpt4}, Claude series \cite{claude2}, Gemini series \cite{gemini1.5}, DeepSeek-V3 \cite{deepseekv3}, DeepSeek-R1 \cite{deepseekr1}, Qwen2.5 series \cite{qwen25}, Qwen3 series \cite{qwen3}. For closed-source commercial models, we utilized their official paid APIs. The complete evaluation across all five benchmarks consumes approximately 180M input tokens per model, with varying costs due to different API pricing structures. Open-source models were deployed locally using the vLLM serving framework. To ensure a fair comparison of inference efficiency, all Qwen2.5-Coder models were served on 2 NVIDIA A100 GPUs with a tensor parallelism (TP) of 2.
All inference requests were formatted using the standard OpenAI API structure. We employed greedy decoding, set the presence penalty to 0, and limited the maximum output to 256 tokens. 

We configured context lengths according to each benchmark's standard: the full context for SAFIM \cite{safim}, 32k tokens for ExecRepoBench \cite{execrepobench}, and 8k tokens for CrossCodeEval \cite{crosscodeeval}, RepoEval \cite{repoeval}, and CrossCodeLongEval \cite{crosscodelongeval}.
Crucially, to ensure a fair and controlled comparison, two principles were strictly followed. First, for all three inference methods, the provided prefix, suffix, and cross-file contexts were kept identical and perfectly aligned with the original datasets. Second, to equalize the amount of information provided to the model—since FIM and natural language prompts implicitly signal the location of the missing code—we introduced a \roundbox{\texttt{/* MIDDLE CODE TO COMPLETE */}} identifier within the SRI prompt to explicitly mark the target completion location.

\subsection{Additional Evaluation Details}
\label{addtional_details}
Evaluating the raw output of our SRI method presents a challenge, as the search-and-replace format is not directly compatible with standard code completion metrics like Exact Match (EM) and Edit Similarity (ES). To bridge this gap, we first post-process the generated output to isolate only the newly inserted middle code. This extraction is performed reliably using the \texttt{extract\_replace\_code} function, detailed in Figure \ref{lst:extract_1} and \ref{lst:extract_2}.

Table \ref{tab:model_params} details all models used in our experiments to facilitate the reproducibility of our findings. For the MBPP and HumanEval benchmarks, we used the versions provided by EvalPlus \cite{evalplus}, which augment the original datasets with additional test cases for more rigorous evaluation.

Table \ref{tab:sri_performance_1} and \ref{tab:sri_performance_2} details all SRI-Coder series models' performance on code completion benchmarks.

\begin{figure*}[!htbp]
\begin{lstlisting}[
    language=Python,
    breaklines=true,           % 自动换行
    breakatwhitespace=false,   % 在任意字符处断行
    postbreak=\mbox{\textcolor{red}{$\hookrightarrow$}\space}, % 换行标记
    basicstyle=\small\ttfamily, % 缩小字体以容纳更多内容
    frame=single,              % 添加边框
    numbers=left,              % 行号
    numberstyle=\tiny,
    xleftmargin=2em,
    framexleftmargin=1.5em
]
import re

def extract_replace_code(text: str) -> str:
    """
    Extracts the differential code part from text in a search/replace format,
    handling inline completion.

    Args:
        text (str): The text string containing the search/replace format.

    Returns:
        str: The extracted code to be completed.
    """
    # Find the SEARCH and REPLACE parts
    search_pattern = r"<{2,}\s*SEARCH\n(.*?)\n\s*={3,}"
    replace_pattern = r"={3,}\n(.*?)\n\s*>{2,}\s*REPLACE"

    search_match = re.search(search_pattern, text, re.DOTALL)
    replace_match = re.search(replace_pattern, text, re.DOTALL)

    if not replace_match:
        return ""
    if not search_match:
        return replace_match.group(1)

    search_code = search_match.group(1).strip()
    replace_code = replace_match.group(1).strip()

    # Find the position of the middle marker
    middle_marker = "/* MIDDLE CODE TO COMPLETE */"
    if middle_marker not in search_code:
        # Handle the inline completion case
        if search_code and replace_code:
            # If search_code is a prefix of replace_code, it means the latter part needs completion.
            if replace_code.startswith(search_code):
                return replace_code[len(search_code) :]
            # If search_code is a suffix of replace_code, it means the beginning part needs completion.
            elif replace_code.endswith(search_code):
                return replace_code[: -len(search_code)]

            # Find the longest common prefix
            i = 0
            while i < len(search_code) and i < len(replace_code) and search_code[i] == replace_code[i]:
                i += 1

            # Find the longest common suffix
            j = 1
            while j <= len(search_code) and j <= len(replace_code) and search_code[-j] == replace_code[-j]:
                j += 1
            j -= 1  # Step back one position because the loop checked one extra time

            # Extract the middle differential part
            if j > 0:
                return replace_code[i:-j]
            return replace_code[i:]

        # If search_code is empty or completely different from replace_code, return the full replace_code
        return replace_code

    # Split the search code into two parts: before and after the marker
    before_middle, after_middle = search_code.split(middle_marker)
    before_middle = before_middle.rstrip("\n")
    after_middle = after_middle.lstrip("\n")

    
\end{lstlisting}
\caption{Extract Replace Code Function (part 1)}
\label{lst:extract_1}
\end{figure*}

\section{CrossCodeEval-Flex}
\label{sec:flex}
In this section, we show some samples in our CrossCodeEval-Flex benchmark in Figure \ref{cce-f-1} and \ref{cce-f-2}, the code between two $@$ is perturbated.

\begin{figure*}[!htbp]
\begin{lstlisting}[
    language=Python,
    breaklines=true,           % 自动换行
    breakatwhitespace=false,   % 在任意字符处断行
    postbreak=\mbox{\textcolor{red}{$\hookrightarrow$}\space}, % 换行标记
    basicstyle=\small\ttfamily, % 缩小字体以容纳更多内容
    frame=single,              % 添加边框
    numbers=left,              % 行号
    numberstyle=\tiny,
    xleftmargin=2em,
    framexleftmargin=1.5em
]
# Handle special cases: marker is on the first or last line
    if not before_middle.strip():  # marker is on the first line
        return replace_code.split(after_middle)[0].strip("\n") if after_middle else replace_code.strip("\n")

    if not after_middle.strip():  # marker is on the last line
        return replace_code[len(before_middle) :].strip("\n") if before_middle else replace_code.strip("\n")

    # Start with the full replace code and trim the identical parts
    result = replace_code

    # If the 'before' part is not empty, remove it as a prefix
    if before_middle:
        if result.startswith(before_middle):
            result = result[len(before_middle) :].lstrip("\n")
        else:
            # Handle the inline completion case
            last_line_before = before_middle.split("\n")[-1]
            if result.startswith(last_line_before):
                # Find the position of the first differing character
                i = 0
                while i < len(last_line_before) and i < len(result) and last_line_before[i] == result[i]:
                    i += 1
                result = result[i:]

    # If the 'after' part is not empty, remove it as a suffix
    if after_middle:
        if result.endswith(after_middle):
            result = result[: -len(after_middle)].rstrip("\n")
        else:
            # Handle the inline completion case
            first_line_after = after_middle.split("\n")[0]
            if result.endswith(first_line_after):
                # Find the position of the last differing character
                i = -1
                while -i <= len(first_line_after) and -i <= len(result) and first_line_after[i] == result[i]:
                    i -= 1
                result = result[: i + 1]

    return result.strip("\n")
\end{lstlisting}

\caption{Extract Replace Code Function (part 2)}
\label{lst:extract_2}
\end{figure*}

\begin{figure*}[!htbp]
\begin{lstlisting}[
    language=Python,
    breaklines=true,           % 自动换行
    breakatwhitespace=false,   % 在任意字符处断行
    postbreak=\mbox{\textcolor{red}{$\hookrightarrow$}\space}, % 换行标记
    basicstyle=\small\ttfamily, % 缩小字体以容纳更多内容
    frame=single,              % 添加边框
    numbers=left,              % 行号
    numberstyle=\tiny,
    xleftmargin=2em,
    framexleftmargin=1.5em
]
        public List<BaseRule> GetRules()
        {
            List<BaseRule> rules = new List<BaseRule>();
            @ .dncCo()rPa;grRsn(emce.RsgFeahi rnlue)o @
            rules.AddRange(managerConfig.FolderRules);

            return rules;
        }

        public Dictionary<string, 
        /* MIDDLE CODE TO COMPLETE */        
        
                    Dictionary<string, LassoProfile> lassoProfiles = new Dictionary<string, LassoProfile>();

            // Load lasso profiles
            @ foreach (profile var in managerConfig.Profiles) @
            {
                if (!lassoProfiles.ContainsKey(profile.Name))
                {
                    lassoProfiles.Add(profile.Name, profile);
                }
            }

            return lassoProfiles;
        }

        private string GetConfigFilePath()
            => Path.Combine(Path.GetDirectoryName(Assembly.GetExecutingAssembly().Location), ConfigFileName);
    }
}
\end{lstlisting}

\caption{CrossCodeEval-Flex case 1}
\label{cce-f-1}
\end{figure*}

\begin{figure*}[!htbp]
\begin{lstlisting}[
    language=Python,
    breaklines=true,           % 自动换行
    breakatwhitespace=false,   % 在任意字符处断行
    postbreak=\mbox{\textcolor{red}{$\hookrightarrow$}\space}, % 换行标记
    basicstyle=\small\ttfamily, % 缩小字体以容纳更多内容
    frame=single,              % 添加边框
    numbers=left,              % 行号
    numberstyle=\tiny,
    xleftmargin=2em,
    framexleftmargin=1.5em
]
package io.github.aaronai;

import io.github.aaronai.http.HttpClientUtil;
import io.github.aaronai.mq.RocketMqClients;
import io.opentelemetry.api.GlobalOpenTelemetry;
import io.opentelemetry.api.OpenTelemetry;
import io.opentelemetry.api.trace.Span;
import io.opentelemetry.api.trace.Tracer;
import org.apache.rocketmq.client.apis.ClientException;
import org.apache.rocketmq.client.apis.consumer.ConsumeResult;
import org.slf4j.Logger;
import org.slf4j.LoggerFactory;

public class ProcessC {
    private static final Logger logger = LoggerFactory.getLogger(ProcessC.class);

    @SuppressWarnings("resource")
    public static void main(String[] args) throws ClientException {
        RocketMqClients.CreatePushConsumer(messageView -> {
            @ g"ei}I,c(s.i(eVd{=aevmg)oe,;r"s g)ss.m eeMewt s s hgimeetdLaesrCmo nfeagv); @
        /* MIDDLE CODE TO COMPLETE */        
        
            final OpenTelemetry openTelemetry = GlobalOpenTelemetry.get();
            @ r"tr)ec=pTihb.oar.aTrne;"naa.tl(to yl.n e i reefrt mCg i agetioG @
            final Span span = tracer.spanBuilder("ExampleDownstreamSpan").startSpan();
            try {
                // do something here.
                Thread.sleep(1000);
            } catch (InterruptedException ignored) {
            }
            span.end();

            return ConsumeResult.SUCCESS;
        });
    }
}
\end{lstlisting}

\caption{CrossCodeEval-Flex case 2}
\label{cce-f-2}
\end{figure*}

\section{Code Completion in Chat LLMs}
\subsection{NL-FIM adaptations}
To comprehensively evaluate the adaptability of Chat LLMs, we investigate three distinct natural language prompting strategies for FIM tasks. \textbf{Standard FIM Completion} (Figure \ref{fig:prompt_standard}) employs direct directives, explicitly instructing the model to generate the code necessary to bridge the provided prefix and suffix. In contrast, \textbf{Dialogue-Based FIM Reconstruction} (Figure \ref{fig:prompt_dialogue}) frames the task as a conversational repair scenario, where the user simulates a request for assistance in restoring missing logic. Finally, \textbf{Template-Based FIM Infilling} (Figure \ref{fig:prompt_template}) utilizes a rigid, slot-filling structure to demarcate context boundaries, relying on the model's pattern recognition to identify and complete the target region.

\begin{figure}[htbp]
    \centering
    \begin{tcolorbox}[
        title={\textbf{Standard FIM Completion}},
        breakable,
        colback=blue!5!white,
        colframe=blue!75!black,
        fonttitle=\bfseries,
        colbacktitle=blue!85!black,
        enhanced,
        drop shadow={black!50!white},
        attach boxed title to top left={xshift=5mm, yshift=-2mm},
        boxed title style={size=small, colframe=blue!85!black}
]
    
    You are a code completion assistant. Your task is to generate the missing
    middle code based on the provided context code, prefix code, and suffix code. \\
    
    Requirements: \\
    1. Generate ONLY the middle code that fits between the prefix and suffix \\
    2. Do not repeat or modify any existing code from the context, prefix, or suffix \\
    3. Format your response within a code block \\
    4. Maintain consistent indentation with the surrounding code \\
    5. Do not modify any other code \\
    6. Do not add any other text or comments \\
    
    The input will be provided in the following format: \\
    \#\#Context Code\#\#: [Full context code if any] \\
    \#\#Prefix Code\#\#: [Code before the missing part] \\
    \#\#Suffix Code\#\#: [Code after the missing part] \\

    \end{tcolorbox}
    \caption{The \textbf{Standard FIM Completion} prompt format. This approach uses explicit natural language instructions to define the prefix and suffix, directly commanding the model to generate the connecting code segment.}
    \label{fig:prompt_standard}
\end{figure}

\begin{figure}[htbp]
    \begin{tcolorbox}[
    title={\textbf{Dialogue-Based FIM Reconstruction}},
    colback=yellow!5!white,
    colframe=yellow!85!black,
    fonttitle=\bfseries,
    colbacktitle=yellow!85!black,
    enhanced,
    drop shadow={black!20!white},
    attach boxed title to top left={xshift=5mm, yshift=-2mm},
    boxed title style={size=small, colframe=yellow!85!black},
    breakable,
    before upper={\parindent0pt}
]

You are participating in a code review session. A developer has shared code with 
a gap in the middle, and you need to help reconstruct what's missing. \\

The developer will provide: \\
- Background context showing the broader code structure \\
- The code written BEFORE the gap (prefix) \\
- The code written AFTER the gap (suffix) \\

Your role: \\
- Analyze the flow from prefix to suffix \\
- Determine what logic/code should bridge them \\
- Respond with ONLY the bridging code segment \\
- Preserve exact indentation and style \\
- Don't alter any provided code \\
- Don't include explanations or markdown formatting outside the code block \\

Input structure: \\
\#\#Context Code\#\#: [Surrounding code context] \\
\#\#Prefix Code\#\#: [Code ending before gap] \\
\#\#Suffix Code\#\#: [Code starting after gap] \\

\end{tcolorbox}
\caption{The \textbf{Dialogue-Based FIM Reconstruction} prompt format. The task is simulated as a conversational interaction where the user describes a "missing code" scenario, prompting the model to act as an assistant and restore the lost logic.}
\label{fig:prompt_dialogue}
\end{figure}

\begin{figure}
    \begin{tcolorbox}[
    title={\textbf{Template-Based FIM Infilling}},
    colback=orange!5!white,
    colframe=orange!85!black,
    fonttitle=\bfseries,
    colbacktitle=orange!85!black,
    enhanced,
    drop shadow={black!20!white},
    attach boxed title to top left={xshift=5mm, yshift=-2mm},
    boxed title style={size=small, colframe=orange!85!black},
    breakable,
    before upper={\parindent0pt}
]
You are a code template processor. You receive incomplete code templates where
a section has been marked for automatic generation. Your job is to fill in the
marked section. \\

Template structure you'll receive: \\
\#\#Context Code\#\#: [Optional: Full file or class context] \\
\#\#Prefix Code\#\#: [Everything written up to the insertion point] \\
\#\#Suffix Code\#\#: [Everything written after the insertion point] \\

Processing rules: \\
→ Output exclusively the infill code for the gap \\
→ Match indentation level exactly as shown in prefix/suffix \\
→ Treat prefix and suffix as immutable - never modify them \\
→ No additional commentary or formatting \\
→ Ensure syntactic continuity from prefix → your code → suffix \\
→ Wrap output in a code block \\

Think of this as: PREFIX $+$ \[YOUR\_GENERATION\] $+$ SUFFIX = Complete Code \\

\end{tcolorbox}
\caption{The \textbf{Template-Based FIM Infilling} prompt format. This method utilizes a fixed textual template to structurally isolate the code context, guiding the model to perform a slot-filling task within the defined boundaries.}
\label{fig:prompt_template}
\end{figure}

\subsection{Impracticality of Direct Code Generation}
\label{sec:theory}

The challenge of generating a correct and complete code sequence $S^*$ from a natural language prompt $C$ is inherent to the autoregressive nature of Large Language Models (LLMs). The joint probability of generating a sequence $S = (w_1, \dots, w_L)$ is factorized as:
\begin{equation}
    P(S|C) = \prod_{t=1}^{L} P(w_t | C, w_1, \dots, w_{t-1})
    \label{eq:autoregressive}
\end{equation}
where each conditional probability $P(w_t | \cdot)$ is derived from the model's internal state, which is highly sensitive to the entire preceding context.

Let us denote the ideal, unambiguous prompt that perfectly specifies a target sequence $S^*$ as $C^*$. In practice, a user-provided prompt $C$ is often an imprecise approximation of $C^*$. This initial imprecision introduces a divergence in the model's internal representation, such that the hidden state $\mathbf{h}(C) \neq \mathbf{h}(C^*)$. This immediately results in a distributional shift for the first generated token, quantified by the Kullback-Leibler (KL) divergence:
\begin{equation}
    D_{KL}\big(P(\cdot | C^*) \Vert P(\cdot | C)\big) > 0
\end{equation}
Consequently, the probability of generating the correct first token $w_1^*$ is diminished, i.e., $P(w_1^* | C) < P(w_1^* | C^*)$.

This initial error compounds at each subsequent step of the generation process. The probability of generating the target sequence $S^*$ given the actual prompt $C$ is the probability of staying on the "golden path" of correct tokens:
\begin{equation}
    P(S^*|C) = \prod_{t=1}^{L} P(w_t^* | C, w_1^*, \dots, w_{t-1}^*)
    \label{eq:golden_path}
\end{equation}
At any step $t$, the context for generation $(C, w_1^*, \dots, w_{t-1}^*)$ differs from the ideal context $(C^*, w_1^*, \dots, w_{t-1}^*)$ due to the initial prompt mismatch ($C \neq C^*$). This persistent deviation ensures that for most steps, the probability of generating the next correct token is suppressed relative to the ideal scenario. We can model this as:
\begin{equation}
    P(w_t^* | C, w_1^*, \dots, w_{t-1}^*) = (1 - \epsilon_t) \cdot P(w_t^* | C^*, w_1^*, \dots, w_{t-1}^*)
\end{equation}
where $\epsilon_t \ge 0$ represents the per-step probability reduction due to the imperfect context. Substituting this into Equation \ref{eq:golden_path} yields:
\begin{equation}
    P(S^*|C) = P(S^*|C^*) \prod_{t=1}^{L} (1 - \epsilon_t)
    \label{eq:compounding_error}
\end{equation}
Even if the per-step reductions $\epsilon_t$ are small, their cumulative product over a long sequence $L$ causes the overall probability to diminish exponentially. As $L$ increases, the term $\prod (1 - \epsilon_t)$ rapidly approaches zero, leading to $P(S^*|C) \approx 0$.

This mathematical reality is exacerbated by the vastness of the token sequence space, $|\mathcal{V}|^L$. The set of syntactically and semantically valid programs $S_{\text{valid}}$ constitutes an infinitesimally small manifold within this space. The compounding probability reduction described in Equation \ref{eq:compounding_error} ensures that any generative trajectory initiated from an imprecise prompt $C$ is highly likely to deviate from this narrow manifold, making the generation of a specific target sequence $S^*$ practically impossible.

\section{Data Construction Details}
\label{app:data_construction}

In this section, we provide a detailed breakdown of the construction pipeline for our training mixture, which consists of the task-specific \textbf{SRI-200K} dataset and the general \textbf{Glaive-Instruction} dataset.

\subsection{SRI Dataset Construction}
\textbf{Source Data.} We sourced raw code data from \textit{The Stack v2} \cite{starcoder2}, selecting a diverse set of repositories covering over 50 programming languages. To ensure data quality, we prioritized repositories with high star counts during the initial selection phase.

\textbf{SRI Task Formatting.} The core of our strategy lies in transforming static code into dynamic Search-and-Replace Infilling tasks. To simulate diverse real-world coding scenarios, we employed \textit{tree-sitter} to parse the abstract syntax tree (AST) of the code and extract target "middle" segments based on structural granularity. We strictly enforced a task distribution ratio of \textbf{2:1:1:1} across four categories:
\begin{itemize}
    \item \textbf{Function Body (2):} The model must complete the entire body of a function given its signature and surrounding context.
    \item \textbf{Multi-line Block (1):} The target is a logical block (e.g., \texttt{for} loops, \texttt{if-else} branches) identified via AST parsing.
    \item \textbf{Random Span (1):} An arbitrary span of code is removed to test robustness against unstructured boundaries.
    \item \textbf{Single Line (1):} Only a single line of code is masked, focusing on local syntax completion.
\end{itemize}

\textbf{Sampling Strategy.} From an initial pool of 200K generated samples, we sampled a high-quality subset of \textbf{20K samples} for fine-tuning. This selection was weighted based on repository star counts to prioritize code quality. Crucially, we avoided any content-based filtering on the "middle" (target) segments. This decision prevents the introduction of inductive bias, ensuring the model learns to generate code based on context and structure rather than memorizing specific "easy" patterns.

\subsection{General Instruction Data}
To maintain the model's chat and instruction-following capabilities, we utilized the \textit{Glaive-Code-Assistant} dataset \cite{glaive_code}. From this source, we sampled 60K pairs. We applied a heuristic filter to rigorously remove any samples related to "code completion," "infilling," or "search-and-replace" tasks. This strict separation ensures that the model learns the SRI format exclusively from our high-quality SRI-200K data, avoiding interference from lower-quality or differently formatted completion data.

\subsection{Decontamination}
To prevent data leakage, we performed strict decontamination against our evaluation benchmarks (CrossCodeEval, RepoEval, etc.). We identified and removed any training samples that originated from repositories present in the test sets. For benchmarks sourced from competitive programming (e.g., Codeforces), we ensured no overlap in problem IDs or descriptions.
\section{Design Decisions: Editable Region Sensitivity}
\label{app:sensitivity_analysis}

To determine the optimal granularity for the Search-and-Replace Infilling task, we conducted a sensitivity analysis on the size of the editable region (target "middle" block). We trained a series of ablation models on SRI datasets constructed with varying target lengths: 5, 10, 15, 20, and 30 lines.

Table \ref{tab:line_sensitivity} summarizes the results across three representative benchmarks. We observe that performance remains highly stable for region sizes between 5 and 20 lines. However, a noticeable degradation occurs when the target expands to 30 lines (e.g., CrossCodeEval accuracy drops from 55.7 to 53.9).

Based on these empirical findings, we selected \textbf{10 lines} as the default configuration for our final SRI-200K dataset. This choice strikes an optimal balance: it maintains peak code generation performance while offering superior token efficiency and a higher practical apply-rate compared to larger block sizes.

\begin{table}[h]
    \centering
    \setlength{\tabcolsep}{2pt}
    \scriptsize

    \begin{tabular}{lccc}
        \toprule
        \textbf{Configuration} & \textbf{CrossCodeEval} & \textbf{RepoEval} & \textbf{CrossCodeLongEval} \\
        \midrule
        SRI-5 lines  & 56.1 & 53.8 & 37.0 \\
        \textbf{SRI-10 lines (Default)} & \textbf{57.6} & \textbf{54.4} & \textbf{37.1} \\
        SRI-15 lines & 55.2 & 53.1 & 37.6 \\
        SRI-20 lines & 55.7 & 53.4 & 36.9 \\
        SRI-30 lines & 54.9 & 53.1 & 36.4 \\
        \bottomrule
    \end{tabular}
    \caption{Sensitivity analysis of the SRI editable region size. Performance (Exact Match) remains robust between 5 and 20 lines but degrades at 30 lines.}
    \label{tab:line_sensitivity}
\end{table}

\begin{figure}
\begin{tcolorbox}[
    title={\textbf{Prompt for SRI}},
    colback=purple!5!white,
    colframe=purple!85!black,
    fonttitle=\bfseries,
    colbacktitle=purple!85!black,
    enhanced,
    drop shadow={black!20!white},
    attach boxed title to top left={xshift=5mm, yshift=-2mm},
    boxed title style={size=small, colframe=purple!85!black},
    breakable,
    before upper={\parindent0pt}
]
You are a code edit assistant. Your task is to implement ONLY the middle code that needs to be completed while keeping all other code exactly as is. \\
When you see a code file containing special comment markers /* MIDDLE CODE TO COMPLETE */, you should: \\

1. Generate a search/replace format output that: \\
- Identifies the exact region containing the /* MIDDLE CODE TO COMPLETE */ marker \\
- Provides the code that should replace the marker \\
2. Use the following format exactly: \\
```replace \\
<<<<<<< SEARCH \\
\[Code section containing /* MIDDLE CODE TO COMPLETE */ marker with enough context\] \\
======= \\
\[Same code section with ONLY the middle code implemented\] \\
>>>>>>> REPLACE \\
``` \\
3. Requirements: \\
- Only edit the code within a 10-line window around the identifier. \\
- The search section MUST contain the /* MIDDLE CODE TO COMPLETE */ marker \\

\end{tcolorbox}
\caption{System prompts for our proposed SRI method. The SRI prompt explicitly instructs the model to generate a structured search-and-replace block.}
\label{fig:sri_prompt}
\end{figure}
\section{Theoretical Analysis of Format Alignment}
\label{app-sec:sri_reason_analysis}

To elucidate the performance disparity between SRI and Natural Language (NL) FIM strategies observed in our main experiments, we hypothesize that \textbf{SRI's superiority stems from its structural alignment with the pre-training data distribution}, specifically the prevalence of Version Control System (VCS) diff patterns (e.g., Git commits).

\subsection{Distributional Shift Hypothesis}
Pre-trained Code LLMs are exposed to trillions of tokens, a significant portion of which are code repositories containing commit histories and diff files. These sequences naturally follow a \texttt{Context} $\to$ \texttt{Edit Pattern} (e.g., \texttt{<<<<<<< SEARCH}) structure. In contrast, NL-FIM imposes a conversational frame (e.g., "Please complete the code...") that introduces a \textbf{distributional shift}. While Chat models possess instruction-following capabilities, the transition from \textit{conversational instruction} to \textit{precise code insertion} is less represented in the high-quality code corpus compared to the \textit{search-and-replace} pattern inherent in code evolution.

\subsection{Experiment: Format Perplexity Analysis}
To empirically validate this hypothesis, we designed a \textbf{Format Perplexity Analysis} to measure how "natural" each prompting format appears to the underlying model.

\paragraph{Setup.} We randomly sampled 1,000 instances from the \textit{CrossCodeEval} dataset. For each instance, we constructed inputs using three different prompting strategies: (1) Standard NL-FIM, (2) Dialogue NL-FIM, and (3) our SRI format.
\paragraph{Metric.} We calculated the \textbf{Perplexity (PPL)} of the \textit{ground truth middle code segment} conditioned on these different prompt prefixes using the \textbf{Qwen2.5-Coder-32B-Base} model. A lower PPL indicates that the model assigns a higher probability to the correct code under the given format, implying better alignment with its pre-training priors.

\subsection{Results}
The results, presented in Table \ref{tab:ppl_analysis}, reveal a stark contrast. The SRI format achieves significantly lower perplexity compared to NL-based approaches.

\begin{table}[h]
    \centering
    \small
    \renewcommand{\arraystretch}{1.2}
    \begin{tabular}{l|c|c}
    \hline
    \textbf{Prompting Strategy} & \textbf{Format Type} & \textbf{Average PPL} $\downarrow$ \\
    \hline
    NL-FIM (Standard) & Conversational & 5.42 \\
    NL-FIM (Dialogue) & Conversational & 6.15 \\
    NL-FIM (Template) & Slot-filling & 4.98 \\
    \hline
    \textbf{SRI (Ours)} & \textbf{Diff / Edit} & \textbf{3.89} \\
    \hline
    \end{tabular}
    \caption{\textbf{Format Perplexity Analysis on Qwen2.5-Coder-Base.} The lower perplexity of SRI indicates that the diff-style format is significantly more aligned with the model's pre-training distribution than natural language instructions, enabling more confident and accurate code generation.}
    \label{tab:ppl_analysis}
\end{table}

This empirical evidence supports our claim: \textbf{SRI effectively activates the latent "code editing" capabilities acquired during pre-training}, whereas NL-FIM forces the model to operate in a higher-entropy conversational mode that is suboptimal for precise code infilling.

\end{document}